\documentclass[longauth]{aa} 
\usepackage{natbib}
\usepackage{graphicx}
\usepackage{txfonts}
\begin{document}
\title{The cosmic star formation rate evolution from $z=5$ to $z=0$\\
  from the VIMOS VLT Deep Survey\thanks{Based on observations
    collected at the European Southern Observatory Very Large
    Telescope, Paranal, Chile, program 070.A-9007(A), and on data
    obtained at the Canada-France-Hawaii Telescope, operated by the
    Institut National des Sciences de l'Univers of the Centre National
    de la Recherche Scientifique of France, the National Research
    Council of Canada, and the University of Hawaii.}}
\author{L. Tresse\inst{1}
\and O.Ilbert\inst{1,2} 
\and E.~Zucca\inst{3}
\and G.~Zamorani\inst{3} 
\and S.~Bardelli\inst{3}
\and S.~Arnouts\inst{1}
\and S.~Paltani\inst{4,5}
\and L.~Pozzetti\inst{3} 
\and D.~Bottini\inst{6}
\and B.~Garilli\inst{6}
\and V.~Le~Brun\inst{1}
\and O.~Le~F\`evre\inst{1}
\and D.~Maccagni\inst{6}
\and J.-P.~Picat\inst{7}
\and R. Scaramella\inst{8,9}
\and M.~Scodeggio\inst{6}
\and G.~Vettolani\inst{8}
\and A.~Zanichelli\inst{8}
\and C.~Adami\inst{1}
\and M.~Arnaboldi\inst{9}
\and M.~Bolzonella\inst{3} 
\and A.~Cappi\inst{3}
\and S.~Charlot\inst{10}
\and P.~Ciliegi\inst{3}  
\and T.~Contini\inst{7}
\and S.~Foucaud\inst{11}
\and P.~Franzetti\inst{6}
\and I.~Gavignaud \inst{12}
\and L.~Guzzo\inst{13}
\and A.~Iovino\inst{13}
\and H.J.~McCracken\inst{10,14}
\and B.~Marano\inst{15}  
\and C.~Marinoni\inst{1,16}
\and A.~Mazure\inst{1}
\and B.~Meneux\inst{6,13}
\and R.~Merighi\inst{3} 
\and R.~Pell\`o\inst{7}
\and A.~Pollo\inst{1,17}
\and M.~Radovich\inst{9}
\and M.~Bondi\inst{8}
\and A.~Bongiorno\inst{15}
\and G.~Busarello\inst{9}
\and O.~Cucciati\inst{13,17}  
\and F. Lamareille\inst{7}
\and G.~Mathez\inst{7}
\and Y.~Mellier\inst{10,14}
\and P.~Merluzzi\inst{9}
\and V.~Ripepi\inst{9}
}
\offprints{Laurence Tresse}
\institute{
Laboratoire d'Astrophysique de Marseille (UMR 6110), CNRS-Universit\'e de Provence, BP8, 13376 Marseille Cedex 12, France \\
\email{Laurence.Tresse@oamp.fr}
\and Institute for Astronomy, 2680 Woodlawn Dr., University of Hawaii, Honolulu, Hawaii 96822, USA
\and INAF-Osservatorio Astronomico di Bologna, via Ranzani 1, 40127 Bologna, Italy 
\and Integral Science Data Centre, ch.d'\'Ecogia 16, 1290 Versoix, Switzerland
\and Geneva Observatory, ch. de Maillettes 51, 1290, Sauverny, Switzerland
\and INAF-IASF, via Bassini 15, 20133 Milano, Italy  
\and Laboratoire d'Astrophysique de l'Observatoire Midi-Pyr\'en\'ees (UMR 5572), CNRS-Universit\'e Paul Sabatier, 14 avenue E. Belin, 31400 Toulouse, France 
\and INAF-IRA, via Gobetti 101, 40129 Bologna, Italy 
\and INAF-Osservatorio Astronomico di Roma, via di Frascati 33, 00040 Monte Porzio Catone, Italy
\and INAF-Osservatorio Astronomico di Capodimonte, via Moiariello 16, 80131 Napoli, Italy 
\and Institut d'Astrophysique de Paris (UMR 7095), 98bis boulevard Arago, 75014 Paris, France 
\and School of Physics \& Astronomy, University of Nottingham, University Park, Nottingham NG72RD, UK
\and Astrophysical Institute Potsdam, An der Sternwarte 16, 14482 Potsdam, Germany
\and INAF-Osservatorio Astronomico di Brera, via Brera 28, 20121 Milano, Italy 
\and Observatoire de Paris-LERMA, 61 avenue de l'Observatoire, 75014 Paris, France 
\and Universit\`a di Bologna, Dipartimento di Astronomia, via Ranzani 1, 40127 Bologna, Italy 
\and Centre de Physique Th\'eorique (UMR 6207), CNRS-Universit\'e de Provence, 13288 Marseille, France
\and Astronomical Observatory of the Jagiellonian University, ul Orla 171, 30-244 Krak\'ow, Poland
\and Universit\`a di Milano-Bicocca, Dipartemento di Fisica, Piazza delle Science 3, 2016 Milano, Italy
}
\date{Received August 1, 2006; accepted April 26, 2007}
\abstract
{The~VIMOS~VLT~Deep~Survey~(VVDS)~was~undertaken to map the
  evolution of galaxies, large scale structures, and active galaxy
  nuclei from the redshift spectroscopic measurements of
  $\sim$10$^{5}$~objects down to an apparent magnitude $I_{AB} = 24$,
  in combination with a multi-wavelength acquisition
  for~radio,~infrared,~optical,~ultraviolet, and~X-rays~data.}
{We~present~the~evolution~of~the~comoving~star~formation~rate (SFR)~density~in
  the redshift range $0 < z < 5$ using the first epoch data release of
  the VVDS, that is 11564 spectra over 2200~arcmin$^2$ in two fields
  of view, the VVDS-0226-04 and the VVDS-CDFS-0332-27, and the
  cosmological parameters ($\Omega_\mathrm{M}$, $\Omega_{\Lambda}$,
  $h$)~=~(0.3, 0.7, 0.7).}
{We study the multi-wavelength  non dust-corrected luminosity
  densities at~$0 < z < 2$ from the rest-frame far ultraviolet to the
  optical passbands, and the rest-frame 1500~\AA\
  luminosity~functions~and~densities~at~$2.7 < z < 5$.}
{They~evolve from $z=1.2$ to $z=0.05$ according to $(1+z)^{x}$ with $x
  = 2.05, 1.94, 1.92, 1.14, 0.73, 0.42$, and 0.30 in the FUV-1500,
  NUV-2800, U-3600, B-4400, V-5500, R-6500, and I-7900 passbands,
  respectively. From~$z=1.2$~to~$z=0.2$ the $B$-band density for the
  irregular-like galaxies decreases markedly by a~factor~$3.5$ while
  it increases by a~factor~$1.7$ for the elliptical-like galaxies. We
  identify several SFR periods; from $z=5$~to~$3.4$ the FUV-band
  density increases by at most 0.5~dex, from $z=3.4$~to~$1.2$ it
  decreases by 0.08~dex, from $z=1.2$ to $z=0.05$ it declines steadily
  by 0.6~dex.  For the most luminous $M_{AB}(1500~\AA)<-21$ galaxies
  the FUV-band density drops by 2~dex from $z=3.9$ to $z=1.2$, and for
  the intermediate $-21<M_{AB}(1500~\AA)<-20$ galaxies it drops by
  2~dex from $z=0.2$ to $z=0$.  Comparing with dust corrected surveys,
  at $0.4 \la z \la 2$ the FUV seems obscured by a constant factor of
  $\sim1.8-2$ mag, while at $z<0.5$ it seems progressively less obscured
  by up to $\sim0.9-1$ mag when the dust-deficient early-type population
  is increasingly dominating the $B$-band density.}
{The~VVDS~results~agree~with~a~downsizing~picture~where~the~most~luminous~sources~cease~to~efficiently~produce~new
  stars 12 Gyrs ago (at $z\simeq 4$), while intermediate luminosity
  sources keep producing stars until 2.5 Gyrs ago (at $z\simeq 0.2$).
  A modest contribution of dry mergers and morphologies evolving
  towards early-type galaxies might contribute to increase the number
  density of the bright early types at $z<1.5$.  Our observed SFR
  density is not in agreement with a continuous smooth decrease since
  $z\sim4$.}
\maketitle
\section{Introduction}
\subsection{The global star-formation history one decade ago}
The history of the comoving space density of the star formation rate
(SFR) is a key study which has undergone a spectacular explosion of
publications since the pioneer work of \citet{madau96} in which galaxy
surveys were used for the first time.  \citet{madau96} derived the
global star formation as a function of redshift combining emissivities
from three distinct surveys: the H$\alpha$-selected UCM survey
\citep{gallego95} at $z < 0.05$, the largest spectroscopic sample at
that time, the $I$-band selected Canada-France Redshift Survey at $0.2
< z < 1$ \citep[CFRS;][]{1996ApJ...460L...1L}, and their own galaxy
sample at $2 < z < 4.5$ using the Lyman continuum break
colour-selection technique \citep{steidel96} applied on the deep
optical imaging, the Hubble Deep Field (HDF) survey.  Nevertheless
high-$z$ sources were lower limits, and incompleteness corrections and
uniform limiting magnitude were required.

The SFR density is usually derived from the mean luminosity density,
or emissivity, that is $\cal{L} = \int^{\infty}_\mathrm{0}
\phi(\mathrm{L}) \mathrm{L} \mathrm{dL}$. At first sight the SFR
density appears a simple and useful tool to trace back the evolution
of star formation and to link it with the evolution of stellar mass,
but a large spread between different measurements have led to
controversy.  Uncertainties in conversion factors from luminosity to
SFR, and in the amount of dust obscured SFR, coupled with the
different selection criteria of each survey and with the uncertainty
about the shape of the lumino\-si\-ty function, meant that the SFR history
of the Universe was poorly determined, and thus hotly debated.  In the
mid-90's the main questions that arose were: is there a peak of the
cosmic star-formation history at $1.3<z<2.7$? By how much does
interstellar dust attenuate the ionizing flux? Is the evolution so
rapid below $z = 1$? Is the high-redshift dropout population
representative? Does the red galaxy population evolve passively or
not?
\subsection{A wealth of multi-wavelength studies}
Measuring the cosmic star-formation history has advanced rapidly in
the last decade meanwhile one had to translate results using the
standard cold dark matter (CDM) into the $\Lambda$CDM cosmology.  It
adds uncertainties since the cosmological cons\-tant introduces a
redshift dependence to the luminosity function unless one can take
back all the data needed to compute the luminosity function. Still the
inferred larger distances and vo\-lumes result in decreasing
luminosities and densities, and thus in a shallower SFR evolution.

On the one hand, a wealth of multi-wavelength emissivities (far- and
near-ultraviolet, far- and near-infrared, radio, H$\alpha$, etc.) has
been obtained, all of them being more or less directly proportional to
the ionizing ultraviolet stellar spectra at $\lambda < 912$~\AA\
mainly produced by massive, young, short-lived stars (OB stars, $t <$
few x $10^6$ yr). On the other hand, supernovae events and their
by-products (neutrinos and gamma rays) are proportional to the
star-death rate and are also used to probe the SFR.  Nevertheless
these results depend most strongly on stars more massive than 3 solar
mass, and require extrapolation of the initial mass function (IMF) to
lower masses to obtain the SFR for all masses.  Futhermore, the
ultimate desired value is the amount of the interstellar gas mass
transformed into stars, and the correlated gas mass return into the
interstellar medium as stars deplete their initial fuel sources; it
requires assumptions and models for stellar atmospheres and stellar
evolution tracks.

An example of an extensive compilation drawn from the literature of
SFR density measurements at $0 < z < 6$ was done by \citet{hopkins04},
and \citet{hopkins06}, and it led to a cosmic SFR history constrained
to within factors of about $\sim3$. There is now growing evidence that
the evolution has no peak at $z\sim1.5$, however it is still unclear
whether at $z>3$ the evolution flattens, or declines or continues to
increase.  Those remaining uncertainties show the necessity to further
investigate galaxy redshift surveys to constrain galaxy assemblies. In
particular, one needs to explore multi-wavelength datasets over the
same field of view, and to obtain homogeneous datasets over a large
redshift range.

A detailed picture of the SFR history is emerging with the advent of
large, deep redshift surveys, coupled with multi-wavelength ground and
space observations over the {\it same} sky area.  These new
multi-wavelength surveys are providing a better understanding of the
nature and the evolution of the galaxy population \cite[i.e.,
e.g.][and references within]{bell04}. Indeed on one hand, one can
probe the same galaxy population at different wavelengths which gives
insights about stellar masses (near-infrared) and star formation
(far-ultraviolet, far-infrared), possibly coupled with spectroscopic
indexes. On the other hand, one can measure reliable comoving volumes
with accurate redshifts and intrinsic luminosities of objects which
both are key measurements to estimate luminosity functions.
Futhermore observing in various windows of the electro\-ma\-gne\-tic
spectrum enable to select the largest galaxy sample which includes
sources with specific energy distributions, and detectable at only
some wavelengths.  The accuracy in the measurement of the shape of the
luminosity function, coupled with the stellar mass-to-light ratio of
galaxies, is crucial to estimate the amount of baryons in stars at a
given epoch of the Universe.
\subsection{The detailed study of the VVDS}
The VIsible Multi-Object Spectrograph (VIMOS) installed on the
European Southern Observatory (ESO) Very Large Telescope (VLT) was
built to produce systematic large redshift surveys thanks to its high
multiplex capabilities, e.g. $\sim$550 $R\simeq230$ spectra of sources
observed simultaneously over 218 arcmin$^2$ \citep{lefevre03}.  We
conducted the VIMOS VLT Deep Survey (VVDS), a major multi-wavelength
spectroscopic survey, to investigate the evolution of galaxies, Active
Galaxy Nuclei (AGN), and large-scale
structures\footnote{http://www.oamp.fr/virmos/}.

Here we detail our work in using the $I$-selected VVDS first epoch
data described in \citet{lefevre05a} to study the evolution of the
luminosity density within the redshift range $0<z<5$.  It is part of a
series of papers which analyse different aspects related to the
luminosity function evolution with this data set.  \citet{ilbert05}
describe the global optical luminosity functions over $0<z<2$.
\citet{zucca06} explore the color-type luminosity functions over
$0<z<1.5$.  \citet{ilbert06a} investigate the contribution of
different morphological types to the luminosity functions.
\citet{ilbert06b} analyse the luminosity functions in different
environments over $0<z<1.5$.  \citet{paltani07} analyse in detail the
1700~\AA\ luminosity function at $3<z<4$.  \citet{arnouts05} des\-cribe
the 1500~\AA\ luminosity functions over $0<z<1.3$ using GALEX-VVDS
data. The infrared luminosity functions with K-band data and with
SWIRE-VVDS data, and the luminosity functions for different
spectroscopic-based classes are in preparation.

This paper is organized as follows. In Section~2 we present the data.
In Section~3 we detail the methods used to estimate the comoving
luminosity densities and associated uncertainties. In Section~4 we
present the multi-wavelength global luminosity densities at $0<z<2$
derived in the UBVRI and in the near-UV, far-UV passbands.  In
Section~5 we compare our results to other surveys at $z<2$.  In
Section~6 we investigate the luminosity densities for different galaxy
types at $z<2$.  In Section~7 we measure the rest-frame 1500~\AA\
luminosity functions and densities at $2.7<z<5$ and we compare our
results to other surveys.  In Section~8 we detail the evolution of the
global far-UV luminosity density all the way from $z=5$ to $z=0$, and
we analyse its dependency to the luminosity.  In Section~9 we derive
the history of the star formation rate density since $z=5$ and discuss
the issue of dust obscuration.  Finally in Section~10 we recap our
conclusions about the evolution of the rest-frame luminosity densities
in a well controlled and homogeneous $I$-selected population over the
large redshift range $0<z<5$ as observed by the VVDS. Throughout this
paper we use the AB flux normalization \citep{oke74}.  We adopt the
set ($\Omega_\mathrm{M}$, $\Omega_{\Lambda}$, $h$)~=~(0.3, 0.7, 0.7)
for the cosmological parameters.
\section{Data}
Our studied sample is taken from the $I$-band selected spectroscopic
data of the first epoch observations obtained in two fields of view,
VVDS-0226-04 and VVDS-CDFS (i.e. VVDS-0332-27 in the Chandra Deep
Field-South) and described in \citet{lefevre05a} and
\citet{lefevre04}.  It consists of 11564 spectra and it covers
2200~arcmin$^{2}$ of sky area observed in five optical passbands $U$,
$B$, $V$, $R$, and $I$ \citep{2004A&A...417..839L}.  In this paper we
consider the well-defined selection function of the spectroscopic
targets selected from the VVDS photometric parent catalogue with
apparent magnitudes in the range $17.50\le I_{AB} \le 24.0$.  We do
not use any serendipitous sources observed randomly in the slit other
than the target. Spectroscopic observations were efficiently targeted
u\-sing the VMMPS tool developed by our team for the spectrograph
VIMOS-VLT/ESO \citep[see][for details]{bottini05}.  We used the red
grism (5500 to 9500~\AA) and a resolution of $R=227$.  No
pre-selection has been applied in terms of colors, sizes, photometric
redshifts, or peculiar sources. The VVDS is based on the sole
criterion of a $I_{AB}$ flux limit.

Spectroscopic observations have been automatically processed using the
VIPGI tool that we developed \citep{scodeggio06} and spectroscopic
determination is described in \citet{lefevre05a}. The 1$\sigma$
accuracy of the redshift measurements is estimated at $0.0009$ from
repeated VVDS observations.  We emphasize that we have obtained an
excellent efficiency for determining redshifts at $z < 2$ and at $z >
2.7$.  At $2<z<2.7$ reliable spectral features are difficult to
detect, and thus the efficiency to measure a redshift in this range is
very poor. Observations extending further to the blue or into the
near-IR are required to fill in this gap with more redshifts.  We
therefore present our measurements in the redshift ranges [$0.05-2$]
and [$2.7-5$].  In total our targets have been classified as 7840
galaxies, 751 stars, and 71 quasars with a reliable spectroscopic
identification at a confidence level higher than 81 percent
(corres\-ponding to the {\sc VVDS} quality flags 2, 3, 4, and 9), 1580
spectra with an uncertain spectroscopic identification at a confidence
level within [48-58] percent (corresponding to the {\sc VVDS} quality
flags 1) and 792 spectra not identified (corresponding to the {\sc
  VVDS} quality flags 0) .  There are 7631 (1182), 31 (47) and 178
(271) reliable (uncertain) galaxy redshifts at $0<z<2$, $2<z<2.7$, and
$2.7<z<5$ respectively. In our study, we exclude the quasars, which
are easily identified thanks to the presence of large broad
spectrocopic emission lines. Under the term 'galaxies' we note that we
include any narrow emission line AGN.

As we do not have a measured redshift for every source to a fixed
magnitude limit in the observed field of view, we introduced a
statistical weight, which is a function of apparent magnitude and
redshift and corrects for sources not observed ({\it Target Sampling
  Rate}; TSR) and for sources for which the spectroscopic
identification failed or is uncertain ({\it Spectroscopic Success
  Rate}; SSR). This statistical weight has been applied to each
measured galaxy at $0<z<2$ as described in \citet{ilbert05}. The SSR
was estimated in two ways; u\-sing the photometric redshifts, and 
using the uncertain redshifts with a confidence level within [48-58]
percent and assuming that the failed identifications have the same
redshift distribution. The two SSR estimates were discrepant in the
redshift bin $1.5<z<2$ only, with the former being twice as large as 
the latter.  Using the deeper multi-wavelength observations of the
Canada-France-Hawaii Legacy
Survey\footnote{http://www.cfht.hawaii.edu/Science/CFHLS/} over the
VVDS-0226 field, \citet{ilbert06c} obtained better photometric
redshifts than in \citet{ilbert05}.  And thus we have refined our SSR
estimations. In the redshift bin $1.4<z<2$ the SSR is changed by a
multiplicative factor of 0.50, otherwise the SSR estimations did not
change within the redshift range $0<z<1.4$ with respect to those
presented in \citet{ilbert05}.

Absolute magnitude measurements are optimized accoun\-ting for the full
information given by the multi-band photometric data in a way which
minimizes the dependency on the templates used to fit the observed
colors \citep[see Fig.~A.1 in][]{ilbert05}.  That is, we automatically
choose the observed apparent magnitude which is as close as possible
to the rest-frame band redshifted in the observer frame, so the
dependency to the template is null or the smallest possible. We use the
templates generated with the galaxy evolution model PEGASE.2
\citep{fioc97}.  Finally, we use a sample of galaxies which are
equally visible, that is within a given absolute magnitude range which
depends on the rest-frame wavelength as we describe in the next
section.
\section{Measuring comoving luminosity densities}
\subsection{Definition} 
Comoving luminosity densities, $\cal{L} = \int^{\infty}_\mathrm{0}
\phi(\mathrm{L}) \mathrm{L} \mathrm{dL}$, depend on the shape of the
luminosity function, $\phi(\mathrm{L}) \mathrm{dL}$.  In our present
study, the galaxy luminosity function (LF) follows a \citet{sche76}
function characterized by a luminosity, L$^{*}$, a faint-end slope,
$\alpha$, and a normalization density parameter, $\phi^{*}$, and thus
$\cal{L} = \phi^{*} \mathrm{L}^{*}$~$\Gamma(\alpha+\mathrm{2})$.  The
LF is a fundamental measurement of the statistical properties of the
population of galaxies; it is the distribution of the comoving number
density of galaxies as a function of their intrinsic luminosity at a
given epoch. Despite its simple definition, its estimation requires
careful analyses of the survey strategy, the selection criteria, and
the completeness.

The faint-end slope is often measured in the range $-1<\alpha<-2$;
thus for a non-diverging density of galaxies, the LF must have a
cut-off at faint luminosities. Such a cut-off has not yet been
observed.  This implies a high-space density of low-luminosity
galaxies, but although these galaxies are very numerous, they
contribute little to the mean luminosity density; for instance,
sources fainter than 0.1L$^*$ contribute less than 20 percent to
$\cal{L}$ for $\alpha <-1.3$ (see Fig.~\ref{fig1}).  As we do not
observe the faintest galaxies, we use the Schechter functional form
for the STY estimate \citep{sandage79} to suppose the behavior of the
LF at low luminosities.  As the three Schechter para\-meters are highly
correlated, it is necessary to build the luminosity function over the
largest possible range of luminosities. Indeed, a weak constraint of
the slope may have a strong impact on the determination of L$^{*}$,
which is directly translated into the luminosity density estimation.
Nevertheless, the latter is more robust than the estimation of each
single parameter of the Schechter function alone.
\subsection{Method} 
We estimated the LF parameters, $\alpha$, M$^{*}$ (or L$^{*}$), and
$\phi^{*}$, u\-sing the Algorithm for Luminosity Functions (ALF)
developed within the VVDS consortium. ALF uses the non-parametric
V$_\mathrm{max}$, SWML and C$^{+}$ and the parametric STY luminosity
function estimators \citep[see Appendixes in][and re\-fe\-rences
within]{ilbert05}.  Each estimator presents advantages and drawbacks,
and each one is affected differently  by different visibility limits
for the various galaxy types detected in deep flux-limited surveys.
Galaxies are not equally visible in the same absolute magnitude range
mainly due to the spectral type dependency on the $k$ corrections.
The bias was quantified in \citet{2004MNRAS.351..541}; it affects
the faint-end slope of the global LF which can be over/underestimated
depending on the adopted estimator. When the differences between the
estimators are larger than the statistical uncertainties, it indicates
the presence of a significant bias.  Thus in a given redshift range
our LF parameters are estimated with data restricted to the absolute
magnitude range in which all galaxy types are visible. It enables us
to calculate an unbiased LF slope.

We apply two approaches to derive the luminosity density from the LF.
First, we derive \emph{minimal} comoving luminosity densities in
summing the LFs over the observed luminosities; in this case there is
no assumption made over the bright or faint end of the LF which may be
not observable in the lowest or highest redshift bins.  Secondly, we
derive the \emph{estimated} mean comoving luminosity densities in
summing the LFs over all luminosities.  As we do not observe the
faintest galaxies, the latter estimates are derived by extrapolating
the LF obtained using the STY estimator. This approach is the only way
to compare data through cosmic epochs since we integrate to the same
faintest luminosity, as long as a cut-off at faint luminosities is not
observed.
\subsection{Uncertainties} 
The LF parameters are correlated to each other, and the effect of this
correlation is that the uncertainty in the LF integral requires the
incorporation of the LF parameter error ellipse, in addition to the
Poisson uncertainties typically quoted.

Thus for the two correlated parameters, $\alpha-M^{*}$, the
uncertainty of $\cal{L}$ is derived from the optimal confidence
regions as determined using the STY errors.  The error ellipse implies
that the uncertainty is not given by the squares of the
one-standard-deviation ($\sigma$) errors of $\alpha$ and $M^{*}$ as
done in the case of two individual, normally-distributed parameters
which give an estimated value within a 68 percent confidence interval.
Actually, the correlation increases the errors on the other parameter.
Indeed, the probability that $\alpha$ and M$^*$ {\it simultaneously}
take on values with the one-$\sigma$ likelihood contour is 39 percent
only. We use the likelihood contour corresponding to a 68 percent
confidence interval, that is at 2.3$\sigma$.  Using the number
counts, we derive $\phi^{*}$ for each point of this contour.  Our
final uncertainties correspond to the two points of the error contour
which give the lowest and highest values of $\cal{L}$.

In this procedure none of the two parameters, $\alpha$ or M$^{*}$, has
been fixed. If we fix one of the two parameters, then it decreases the
errors on the other parameter. In this case, errors are usually not
realistic, so we use another method to give uncertainties as follows.
We derive $\mathrm{L}^{*}$~$\Gamma(\alpha+\mathrm{2})$ using the
estimated Schechter parameters with the low, high, and mean values
chosen for the fixed parameter, and with the 1$\sigma$ error on the
other free para\-me\-ter. Using the number counts, we derive $\phi^{*}$
for each point of the single free parameter axis.  In the case of a
fixed parameter, we take the extreme uncertainties given by the
highest (lowest) $\cal{L}$ estimation calculated with the high (low)
value of the fixed parameter, and subtracted to the lowest (highest)
estimation calculated with the mean value.  When we fix either M$^*$
or $\alpha$, we footnote it in our tables of $\cal{L}$ values.
\subsubsection{Fixing  the Schechter parameter M$^{*}$} 
The brightest galaxies are not sampled in the first redshift bin
($0.05<z<0.2$) because of the VVDS bright limit at $I_{AB}=17.5$.
Thus in this redshift range we fix M$^{*}\pm0.05$ to its mean local
value, as found by the SDSS at $z=0.1$ in the U, B, V, R, and I bands
\citep{blanton03}. We gave these values in Section 5.1 and Table~1 of
\citet{ilbert05}.  In the far-UV band we fix M$^{*}\pm0.10$ to the one
derived by \citet{wyder05} from the GALEX-2dFGRS survey at $<z>=0.055$
The resulting STY parameters are with $M^{*}_{\rm fix} = -18.12$ mag,
$\alpha=-1.13$, and $\phi^*=0.00732$ Mpc$^{-3}$.
\begin{figure}[t]
\resizebox{\hsize}{7cm}{\includegraphics{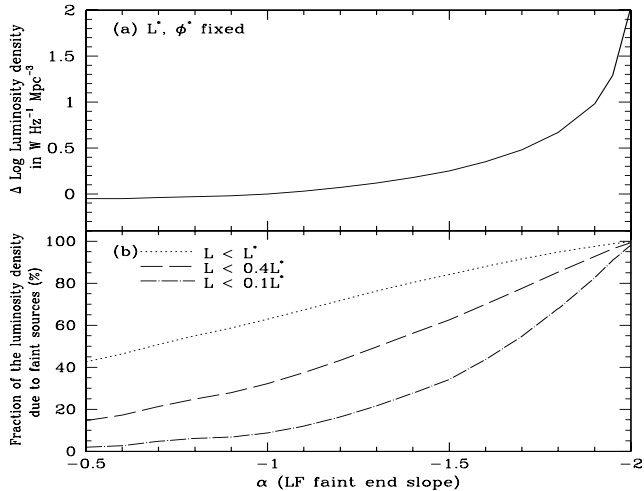}}
\caption{(a). Quantitative change in log of the comoving luminosity
  density, $\cal{L}$, as a function of the faint-end slope, $\alpha$,
  of the LF for fixed L$^*$ and $\phi^*$ values, compared to
  Log~$\cal{L}$ for $\alpha=-1$. (b) Fraction of $\cal{L}$ as a
  function of $\alpha$ due to sources fainter than L$^*$ (dotted
  line), 0.4L$^*$ (dashed line), and 0.1L$^*$ (dot-dashed line).}
\label{fig1}
\end{figure}
\subsubsection{Fixing the Schechter parameter $\alpha$} 
At high redshift ($z>1.2$), we need to fix $\alpha$ because the VVDS
faint limit, $I_{AB}=24$, prevents us from observing luminosities faint
enough to constrain the LF faint-end slope.  One of the most important
things done to obtain a reliable estimate of $\cal{L}$ is to sample the LF
around the knee of the luminosity distribution. If we do not observe
magnitudes fainter than the LF knee, (i.e. $L<L^{*}$), then $\cal{L}$
is underestimated whatever the value given to $\alpha$.  As long as
$\alpha$ is greater than $-1$ the total luminosity density is little
dependent on the slope, as illustrated in Fig.~\ref{fig1}, and is
dominated by the product of $\phi^{*}$ and $L^{*}$. For slopes~$<-1$,
and constant $\phi^{*} \times L^{*}$, $\cal{L}$ increases of 32, 78, 
and 209 percent from $\alpha=-1$ to $-1.3, -1.5$, and $-1.7$,
respectively. Thus uncertainties on $\cal{L}$ can be quite large when
$\alpha < -1.3$.  The low, high, and mean values for a fixed slope are
described in the appropriate sections.
\section{Global luminosity densities at $0<z<2$}
\subsection{The VVDS multi-wavelength data}
The Schechter parameters are those derived in \citet{ilbert05} in the
rest-frame U-3600, B-4400, V-5500, R-6500, and I-7900 passbands from
$z=0.05$ to $z=1$.  In the redshift bins $1.0<z<1.3$ and $1.3<z<2.0$
we were very cautious in fixing the LF faint-end slope to the value of
the slope estimated in the redshift bin $0.8<z<1.0$.  Now, we estimate
the slope in the redshift bin $1.0<z<1.2$, and at $z>1.2$ we use our
result from Fig.~9 in \citet{ilbert05} illustrating the steepening of
the LF faint-end slope with increasing redshift. It corresponds to
$\Delta\alpha =0.3$ between $z\simeq0.1$ and $z\simeq 0.9$, that is an
increase of $0.0375$ at each step of 0.1 in redshift.  We have fixed
$\alpha$ in keeping this increment since there is no reason why it
should suddenly stop increasing.  In particular, for the bin
$1.2<z<1.4$ ($<z>=1.3$), we take the value of $\alpha$ measured in the
bin $1.0<z<1.2$ ($<z>=1.1$) and we increase it by
$2\times0.0375=0.075$.  For the bin $1.4<z<2.0$ ($<z>=1.5$), we take
the value of $\alpha$ in the bin $1.2<z<1.4$ ($<z>=1.3$) and we
increase it by $0.075$ also. In these two redshift bins, we assume
$\pm0.075$ for the error bars of the fixed $\alpha$ values.

We have derived the rest-frame FUV-1500 and NUV-2800 luminosity
densities. While the rest-frame 2800~\AA\ luminosities are sampled by
the observed reference bands from $U$-band to $I$-band from $z=0.3$ to
$z=2.0$, the rest-frame 1500~\AA\ luminosities are sampled only at
$z>1.4$.  Our LFs computed at 1500~\AA\ therefore strongly rely on the
spectral eneygy distribution (SED) fit at longer wavelengths.
However, in \citet{arnouts05} we have already derived the 1500~\AA\
LFs using data from GALEX \citep{martin05}. These data were a
($NUV$-2300+$I$)-selected GALEX-VVDS sample from $z=0.2$ to $z=1.2$
and they were mainly based on a one-to-one identification between
optical images and NUV images. Our present FUV results are 
SED-dependent but derived from a pure $I$-selected sample similar to what 
was done in the other bands. At $0.2<z<0.8$ the $I$-selected VVDS and the
$(NUV+I)$-selected GALEX-VVDS LFs are in agreement within the error
bars. This implies that even though our FUV result is SED dependent,
our SED fitting is globally correct using a fixed $-1.6$ slope at
$0<z<2$. At $0.8<z<1.2$, the two FUV luminosity densities differ
significantly of $\sim$0.2~dex because of a lower $\phi^*$ for the
$(NUV+I)$-selected sample. The weights we adopted in \cite{arnouts05}
were preliminary and slightly underestimated at $0.8<z<1.2$.

\begin{figure}
\resizebox{\hsize}{!}{\includegraphics{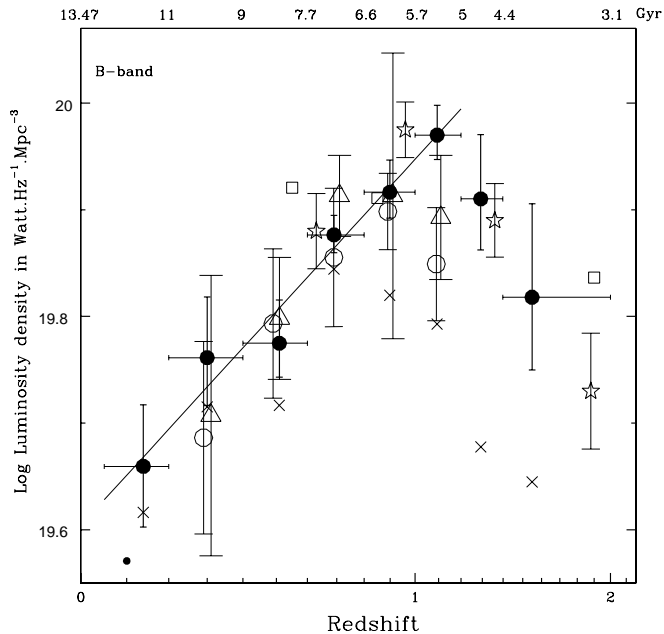}}
\caption{The \emph{estimated} non dust-corrected comoving luminosity
  densities in the rest-frame B-4400 passband as a function of
  $log(1+z)$ and represented with {\it filled circles}. The {\it
    crosses} are the \emph{minimal} comoving densities, derived from
  the observed luminosities without extrapolation neither at the
  bright nor at the faint part of the LF.  Data are displayed at the
  mean redshift of the galaxies within each redshift bin. Error bars
  are derived from the 2.3$\sigma$ error contours of the Schechter
  parameters, except at $z<0.2$ where $M^*$ is fixed and at $z>1.2$
  where $\alpha$ is fixed. The {\it dot} at $z=0.1$ is the SDSS local
  point. The VVDS $B$-band luminosity densities increase as
  $(1+z)^{1.14}$ up to $z=1.1$. Data are listed in Tab.~\ref{table1}.
  The {\it open triangles} and {\it open circles} are the DEEP2 and
  COMBO-17 data displayed for clarity at $-0.05$ and $+0.05$ in
  redshift respectively and taken from Table~2 in \citet{faber05}, the
  {\it open squares} are the HDF data from \citet{poli03}, and the {\it
    open stars} are the FDF data from \citet{gabasch04}.}
\label{fig2} 
\end{figure}

\begin{figure}
\resizebox{\hsize}{!}{\includegraphics{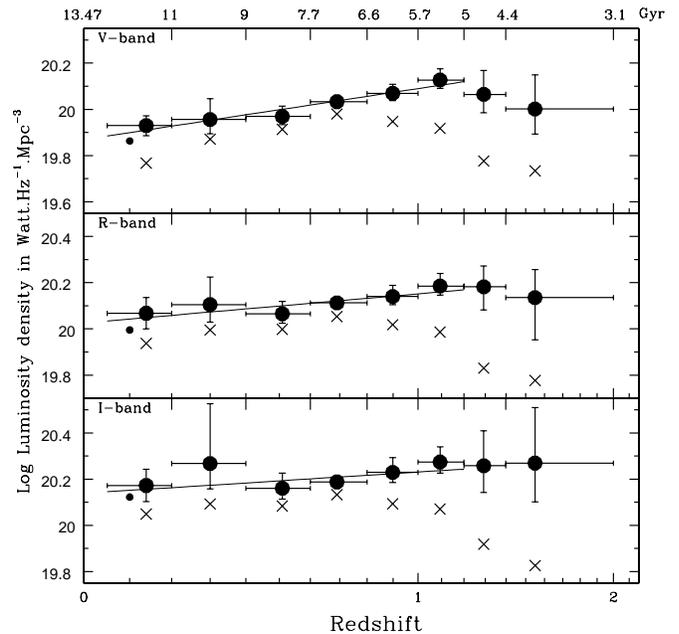}}
\caption{Comoving non dust-corrected luminosity densities in the
  rest-frame V-5500, R-6500, and I-7900 passbands. The {\it filled
    circles}, {\it crosses}, and the {\it dot} are the same as in
  Fig.~\ref{fig2}. The VVDS $V$-, $R$-, and $I$-band densities increase
  as $(1+z)^{0.73}$ , $(1+z)^{0.42}$, and $(1+z)^{0.30}$ up to $z=1.1$
  respectively.}
\label{fig3}
\end{figure}

\begin{figure}
\resizebox{\hsize}{!}{\includegraphics{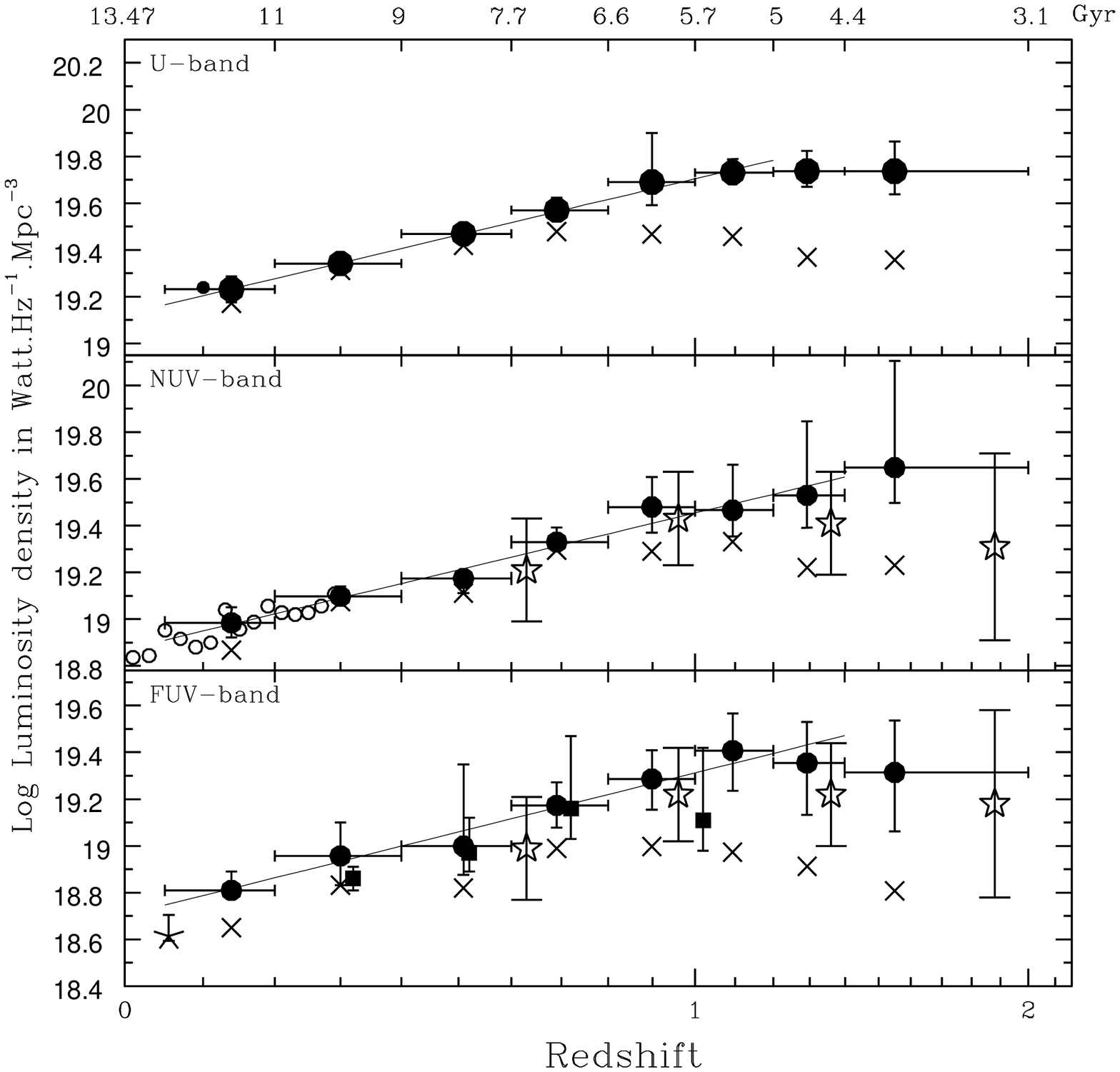}}
\caption{Comoving non dust-corrected luminosity densities in the
  rest-frame U-3600, NUV-2800, and FUV-1500 passbands. The {\it filled
    circles}, {\it crosses}, {\it open stars}, and the {\it dot} come
  from the same data as in Fig.~\ref{fig2}. The VVDS $U$-, $NUV$-, and
  $FUV$-band densities increase as $(1+z)^{1.92}$ , $(1+z)^{1.94}$, and
  $(1+z)^{2.05}$ up to $z=1.1$ respectively. The {\it asterisk} is the
  $FUV$-band local point from the GALEX-2dFGRS survey
  \citep[$<z>=0.055$][]{wyder05}.  The {\it plain squares} are the
  FUV-band data, displayed for clarity at $+0.02$ in redshift, and
  taken from the GALEX-VVDS survey \citep{schiminovich05}. The {\it
    small open circles} are the $NUV$-band points from the SDSS
  \citep{baldry05}.  }
\label{fig4}
\end{figure}

Values of $\cal{L}$ for each passband in various redshift bins are
given in Tab.~\ref{table1}, and they are displayed in
Figs.~\ref{fig2}, ~\ref{fig3}, and ~\ref{fig4} for the B, V\&R\&I, and
U\&NUV\&FUV bands, respectively.
\begin{table*}
  \caption{Comoving multi-wavelength non dust-corrected luminosity densities at $0<z<2$ of the VVDS with the cosmology ($\Omega_\mathrm{M}$, $\Omega_{\lambda}$, $h$)~=~(0.3,0.7,0.7).}
\label{table1}
\begin{center}
\begin{tabular}{ c l c l l l l l l l }
\hline
$<z>^a$ & Redshift range  & $\delta t^b$  &  $\log \cal{L}_\mathrm{1500~\AA}$ & $\log \cal{L}_\mathrm{2800~\AA}$  &  $\log \cal{L}_\mathrm{3600~\AA}$  &  $\log \cal{L}_\mathrm{4400~\AA}$  & $\log \cal{L}_\mathrm{5500~\AA}$ & $\log \cal{L}_\mathrm{6500~\AA}$ & $\log \cal{L}_\mathrm{7900~\AA}$  \\
           &  & &   \multicolumn{7}{c}{W Hz$^{-1}$ Mpc$^{-3}$}               \\
\hline
          &   &               &   \multicolumn{7}{c}{Minimal}             \\
\hline  
$0.14$ & [0.05-0.20] & 2.5  & $18.65$ & $18.87$ &   $19.17$     &     $19.62$   &     $19.77$ &     $19.94$   &      $20.05$ \\
$0.30$ & [0.20-0.40] & 2.0  & $18.83$ & $19.07$ &   $19.31$     &     $19.72$   &     $19.87$ &     $19.99$   &      $20.09$ \\ 
$0.51$ & [0.40-0.60] & 1.3  & $18.82$ & $19.11$ &   $19.42$     &     $19.72$   &     $19.91$ &     $20.00$   &      $20.08$ \\ 
$0.69$ & [0.60-0.80] & 1.1  & $18.99$ & $19.30$ &   $19.48$     &     $19.84$   &     $19.98$ &     $20.05$   &      $20.13$ \\ 
$0.90$ & [0.80-1.00] & 0.9  & $19.00$ & $19.29$ &   $19.47$     &     $19.82$   &     $19.95$ &     $20.02$   &      $20.09$ \\ 
$1.09$ & [1.00-1.20] & 0.7  & $18.97$ & $19.33$ &   $19.46$     &     $19.79$   &     $19.92$ &     $19.99$   &      $20.07$ \\ 
$1.29$ & [1.20-1.40] & 0.6  & $18.91$ & $19.22$ &   $19.37$     &     $19.68$   &     $19.78$ &     $19.83$   &      $19.92$ \\ 
$1.55$ & [1.40-2.00] & 1.3  & $18.81$ & $19.23$ &   $19.36$     &     $19.64$   &     $19.73$ &     $19.78$   &      $19.83$ \\ 
\hline
            &  &             &   \multicolumn{7}{c}{Estimated}             \\
\hline  
$0.14$  & [0.05-0.20]$^c$    & 2.5  & $18.81^{+0.08}_{-0.04}$ & $18.98^{+0.07}_{-0.06}$ &  $19.23^{+0.06}_{-0.06}$      &     $19.66^{+0.06}_{-0.06}$   &    $19.93^{+0.04}_{-0.04}$ &     $20.07^{+0.07}_{-0.07}$   &    $20.17^{+0.07}_{-0.07}$ \\
$0.30$  & [0.20-0.40]$^f$            & 2.0  &  $18.96^{+0.14}_{-0.13}$    & $19.10^{+0.04}_{-0.03}$  & $19.34^{+0.03}_{-0.03}$      &     $19.76^{+0.06}_{-0.04}$   &    $19.96^{+0.09}_{-0.06}$ &     $20.10^{+0.12}_{-0.07}$   &    $20.27^{+0.26}_{-0.11}$ \\ 
$0.51$  & [0.40-0.60]$^{e,f}$     & 1.3  & $19.00^{+0.35}_{-0.12}$    &   $19.17^{+0.02}_{-0.06}$ & $19.47^{+0.03}_{-0.03}$      &     $19.77^{+0.04}_{-0.03}$   &    $19.97^{+0.04}_{-0.04}$ &     $20.07^{+0.05}_{-0.04}$   &    $20.16^{+0.07}_{-0.05}$    \\ 
$0.69$  & [0.60-0.80]$^{e,f}$     & 1.1  & $19.17^{+0.10}_{-0.09}$    &   $19.33^{+0.06}_{-0.03}$ & $19.57^{+0.05}_{-0.04}$      &     $19.87^{+0.02}_{-0.02}$   &    $20.03^{+0.02}_{-0.02}$ &     $20.11^{+0.02}_{-0.02}$   &    $20.19^{+0.03}_{-0.02}$    \\ 
$0.90$  & [0.80-1.00]$^{e,f}$     & 0.9 &  $19.29^{+0.12}_{-0.13}$    &  $19.48^{+0.13}_{-0.11}$ & $19.69^{+0.21}_{-0.10}$      &     $19.92^{+0.03}_{-0.02}$   &    $20.07^{+0.04}_{-0.03}$ &     $20.14^{+0.05}_{-0.04}$   &  $20.23^{+0.06}_{-0.04}$       \\ 
$1.09$  & [1.00-1.20]$^{e,f}$     & 0.7  &  $19.41^{+0.16}_{-0.17}$   &  $19.47^{+0.19}_{-0.11}$ & $19.73^{+0.06}_{-0.05}$      &     $19.97^{+0.03}_{-0.02}$   &    $20.13^{+0.05}_{-0.04}$  &   $20.19^{+0.05}_{-0.04}$   &    $20.27^{+0.07}_{-0.05}$      \\ 
$1.29$  & [1.20-1.40]$^{d,e,f}$  & 0.6  & $19.35^{+0.18}_{-0.22}$    &  $19.53^{+0.32}_{-0.14}$ &  $19.74^{+0.09}_{-0.07}$      &     $19.87^{+0.09}_{-0.04}$   &    $20.06^{+0.10}_{-0.08}$  &   $20.18^{+0.09}_{-0.10}$   &    $20.26^{+0.15}_{-0.12}$      \\ 
$1.55$  & [1.40-2.00]$^{d,e,f}$  & 1.3  &  $19.31^{+0.22}_{-0.25}$   &  $19.65^{+0.46}_{-0.15}$ & $19.74^{+0.13}_{-0.10}$      &     $19.74^{+0.10}_{-0.05}$   &    $20.00^{+0.15}_{-0.11}$  &   $20.14^{+0.12}_{-0.18}$   &    $20.27^{+0.24}_{-0.17}$        \\ 
\hline
\end{tabular}
\end{center}
$^a$Mean redshift of the galaxy redshifts in the quoted redshift range. \\ 
$^b$Elapsed time in Gyr in the redshift bin. \\
$^c$At $0.05<z<0.20$ with  $< z > = 0.14$ the STY LF estimators are derived using a fixed M$^*$ in the U, B, V, R, and I bands taken from the SDSS ($< z >\sim0.1$ \citet{blanton03}; see also Section 5.1 in \citet{ilbert05}), in the FUV-1500 band from the GALEX-2dFGRS survey \citep[$<z>=0.055$][]{wyder05}, and in the NUV-2800 from our value estimated at $0.1<z<0.3$ (i.e. $-18.61$ mag) since there is no local value in the literature. Error bars are derived from the STY estimations at M$^*\pm0.10$ for the FUV and NUV bands, otherwise at M$^*\pm0.05$. \\
$^d$At $1.2<z<2.0$ the STY LF estimators are constrained using a fixed $\alpha$ and error bars are derived from the STY estimations at $\alpha\pm0.0375$ as detailed in Section 4 except for the FUV and NUV bands (see below). \\
$^e$In the NUV the slope is fixed to the value estimated at $0.2<z<0.4$, i.e. $\alpha = -1.32$. Error bars are derived from the STY high estimations assuming an increase of $\alpha$, of $+0.0375$  each 0.1 redshift bin, as observed in the visible (see Section 4), and  from the STY low estimations  assuming $\alpha=-1$ at $0.8<z<2$, and from the minimal value at $0.4<z<0.8$ since in this redshift range the latter is higher than the STY estimate assuming $\alpha=-1$. \\
$^f$In the FUV the slope  is fixed to $\alpha = -1.60$ at $0.2<z<2$ corresponding to the slope found with the GALEX-VVDS sample \citep{arnouts05}. Error bars are derived from the STY estimations assuming $\alpha = -1.20$ and $\alpha = -1.75$, which corresponds to the large range of values found in the literature due to the lack of constraint on the faint end slope of the LF.  
\end{table*}
\subsection{Evolution according the rest-frame passband}
\begin{figure}
\resizebox{\hsize}{!}{\includegraphics{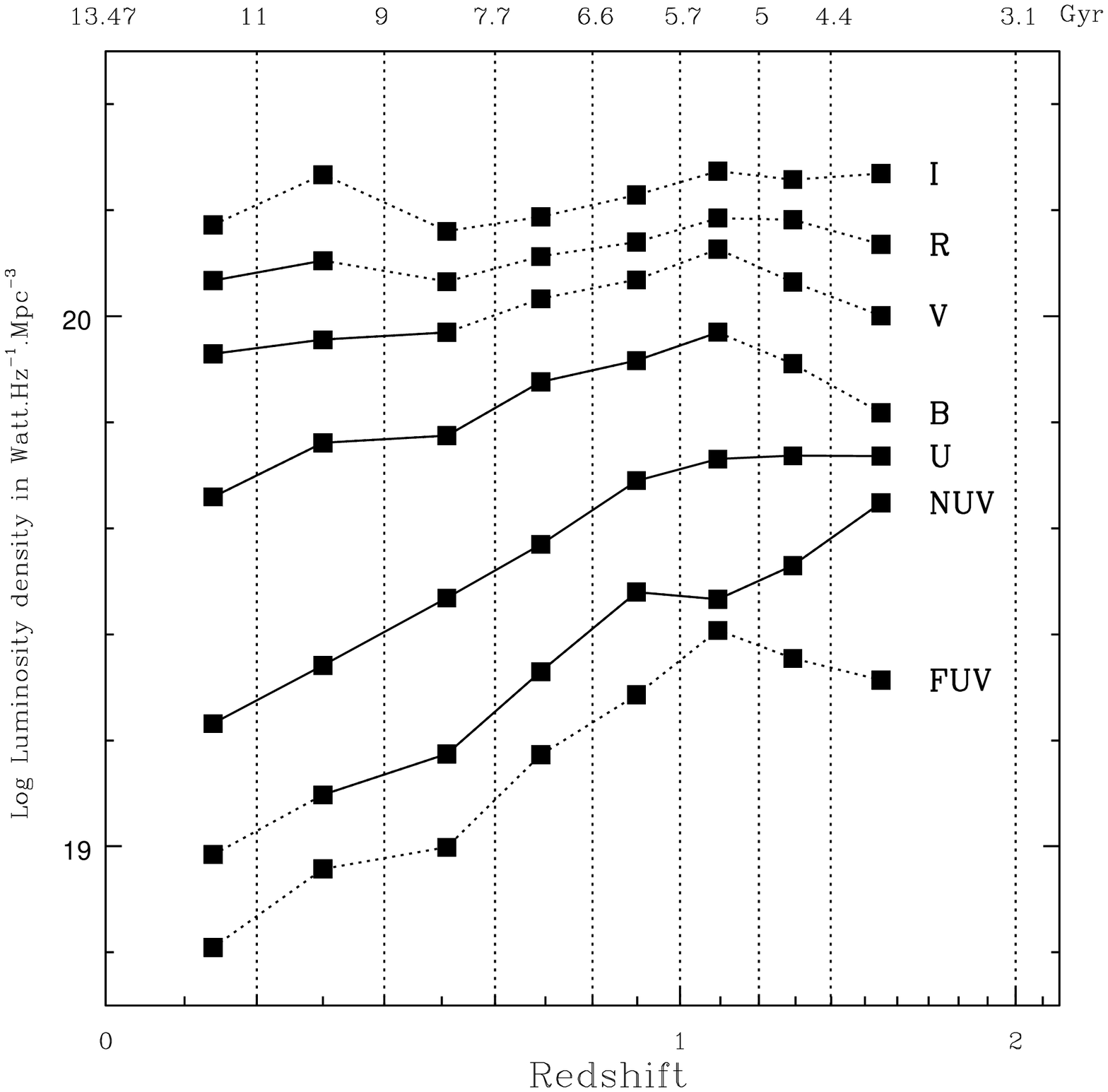}}
\caption{Comoving non dust-corrected VVDS luminosity densities in the
  rest-frame FUV-1500, NUV-2800, U-3600, B-4400, V-5500, R-6500, and
  I-7900 passbands from bottom to top respectively, as displayed in
  Figs.~\ref{fig2}, ~\ref{fig3}, and ~\ref{fig4}. The solid line
  connects points where the rest-frame band is observed in the
  optical.}
\label{fig5}
\end{figure}

The dependence of the emissivity of the global population on the
rest-frame band is noticeable. The non dust-corrected luminosity
densities evolve with redshift over $0.05\le z\le 1.2$, as $\cal{L}
\propto \rm (1+z)^{x}$ with $x=2.05, 1.94, 1.92, 1.14, 0.73, 0.42$,
and $0.30$ in FUV-1500, NUV-2800, U-3600, B-4400, V-5500, R-6500, and
I-7900 passbands, respectively (see Figs.~\ref{fig2}, ~\ref{fig3},
~\ref{fig4}, and ~\ref{fig5}).  There is a clear differential,
wavelength-dependent evolution of the whole po\-pu\-lation.  Indeed, the
average (FUV$-$I) rest-frame color emis\-si\-vi\-ty of the whole galaxy
population becomes four times redder from $z=1.2$ to nowadays.
Futhermore, the data suggest an up-turn in the emissivity evolution at
redder wavelengths than the $I$-band.  Nevertheless a possible up-turn
is likely a selection effect due to the fact that the rest-frame
$I$-band emissivities from an $I$-selected sample could be
underestimated in missing the very red galaxies at $z>1$.

The FUV is related to the formation of young, massive, short-lived,
hot stars, while the NIR is related to long-lived, old stars which
relate closely to the stellar mass of a galaxy.  And thus, over the
last 8.5 Gyrs ($z<1.2$) there has been a substantial decline of the
star formation rate, while the old stellar content shows a smoother
change in terms of emissivity.  In \citet{pozzetti07}, we derive the
stellar masses using a rest-frame K-band sample, and we conclude that
at $z<1$ the stellar mass density increases by a factor of $\sim2.3$.
Since the global rest-frame ultraviolet emissivity continues to
decline, merger events should produce little star formation, either
via minor mergers with e.g. satellite galaxy accretion, or via majors
mergers bet\-ween cold gas-depleted galaxies so there is not enough gas
to efficiently produce new stars.

At $z\sim1.1$ our luminosity densities exhibit a transition in the
evolutionary trend.  In particular, from $z=2$ to $z=1.2$, the $FUV$-,
$B$-, $V$- and $R$-band $\cal{L}$ increase, the $U$- and $I$-band
$\cal{L}$ flatten; and then below $z=1.2$, they all decrease. Only the
NUV $\cal{L}$ follows a continuous evolution since $z=2$.
Nevertheless, error bars at $z>1.2$ are large, and thus the observed
transition is still uncertain.
\section{Comparison with other surveys} 
\subsection{Comparison with the CFRS at $I_{AB} = 22.5$} 
The CFRS is an $I$-selected survey like the VVDS.  The CFRS galaxy
sample consists of 730 $I$-band selected galaxies at $17.5 \le I_{AB}
\le 22.5$, of which 591 (i.e., more than 80 percent) have reliable
redshifts in the range $0 < z < 1.3$ \citep{1995ApJ...455...60L}.
\citet{1996ApJ...460L...1L} estimated the comoving luminosity
densities of the Universe from the CFRS sample in the rest-frame
$NUV$-2800, $B$-4400, and $NIR$-10000 passbands over the redshift range
$0 < z < 1$ with the cosmology ($\Omega_\mathrm{M}$,
$\Omega_{\Lambda}$, $h$)~=~(1, 0, 0.5). Here, we compare with the
$B$-band results which require very little extrapolation from models,
contrary to the NUV or NIR data. From the best estimate of the
$B$-4400 band LF \citep[see][]{1995ApJ...455..108L}, the rest-frame
$B$-band e\-mis\-si\-vi\-ties directly-observed (correspond to our minimal
estimates) and estimated were derived. The CFRS estimated
$\cal{L}_\mathrm{4400~\AA}$ was derived fitting with a Schechter
function the data given by the V$_\mathrm{max}$ LF estimator. The VVDS
estimated $\cal{L}_\mathrm{4400~\AA}$ is derived from the integration
of the LF estimate with $\alpha$, M$^{*}$, and $\phi^{*}$ as determined
with the STY method.  The two fits give similar results as long as
each type of galaxies is visible within the redshift range studied
\citep[see][]{2004MNRAS.351..541}.
\begin{table}
  \caption{Comparison of comoving luminosity densities between $I$-selected surveys, the CFRS and VVDS, with the cosmology ($\Omega_\mathrm{M}$, $\Omega_{\Lambda}$, $h$)~=~(1, 0, 0.5).}
\label{table2}
\begin{center}
\begin{tabular}{c l l l}
\hline
              &   \multicolumn{3}{c}{$\log \cal{L}_\mathrm{4400~\AA}$  in W~Hz$^{-1}$~Mpc$^{-3}$}        \\
\hline
\hline
Redshift      &  CFRS                 &  VVDS              & VVDS  \\
range          &  [17.5-22.5]      &  [17.5-22.5]   & [17.5-24.0]  \\
\hline
                     &     \multicolumn{3}{c}{Minimal}             \\
\hline
$z<0.2$      &             $-$                &        $-$                   &   $19.50\pm0.03$$^a$     \\
 0.20-0.50   &  $19.61\pm0.07$       &   $19.68\pm0.01$    &   $19.67\pm0.01$        \\ 
 0.50-0.75   &  $19.77\pm0.07$       &   $19.80\pm0.02$    &   $19.86\pm0.01$        \\ 
 0.75-1.00   &  $19.74\pm0.07$       &   $19.71\pm0.01$    &   $19.85\pm0.01$        \\ 
\hline
              &     \multicolumn{3}{c}{Estimated}                     \\
\hline
$z<0.2$     &  $19.30\pm0.10^b$   &    $-$                                     &    $19.59^{+0.10}_{-0.10}$$^a$   \\
0.20-0.50   &  $19.63\pm0.07$       &   $19.68^{+0.03}_{-0.03}$    &    $19.70^{+0.04}_{-0.03}$      \\ 
0.50-0.75   &  $19.86\pm0.08$       &   $19.88^{+0.05}_{-0.04}$    &    $19.90^{+0.02}_{-0.02}$       \\ 
0.75-1.00   &  $20.05\pm0.13$       &   $20.06^{+0.72}_{-0.16}$    &    $19.93^{+0.02}_{-0.02}$         \\ 
\hline 
\end{tabular}  
\end{center}
$^a$ With M$^{*}$ fixed to the SDSS $^{0.1}g$-LF estimate from \citet{blanton03} ($B$-band~$\sim\  ^{0.1}g$-band with less than 0.05 mag difference). \\
$^b$ Based on SAPM $b_j$-LF estimate from \citet{1992ApJ...390..338L}.
\end{table}

The upper panel of Fig.~\ref{fig6} compares the CFRS
$\cal{L}_\mathrm{4400~\AA}$ to the VVDS $\cal{L}_\mathrm{4400~\AA}$
with a magnitude cut at $I_{AB} = 22.5$ over the same redshift range,
$0.2 \le z \le 1.0$, and with the same cos\-mo\-lo\-gy adopted in the CFRS
analysis.  The CFRS and the VVDS-[$17.5-22.5$]
$\cal{L}_\mathrm{4400~\AA}$ are very well consistent with each other
(see Tab.~\ref{table2}).  Even though the CFRS contains $\sim4.5$
times fewer galaxies than the VVDS at the same depth, the CFRS error
bars are smaller than the VVDS error bars. This is due to different
procedures used to estimate the uncertainties in the two surveys. The
CFRS 'ad-hoc' uncertainty procedure is described in
\citet{1996ApJ...460L...1L}, and ours are derived from the
$\alpha-M^{*}$ error contour of 68 percent confidence level of our
STY estimate.  In both surveys, the minimal and the estimated $B$-band
emissivities become more discrepant as the redshift increases. This is
due to the combination of brighter limiting luminosities sampled at
higher redshifts, and to a lesser extent, a steeper faint-end slope of
the rest-frame $B$-band LF.  The CFRS $0.2 < z <1.0$ values are thus
found to be very reliable up to $I_{AB} =22.5$.
\subsection{From the CFRS to the 1.5 mag deeper VVDS} 
\begin{figure}[t]
\resizebox{\hsize}{10cm}{\includegraphics{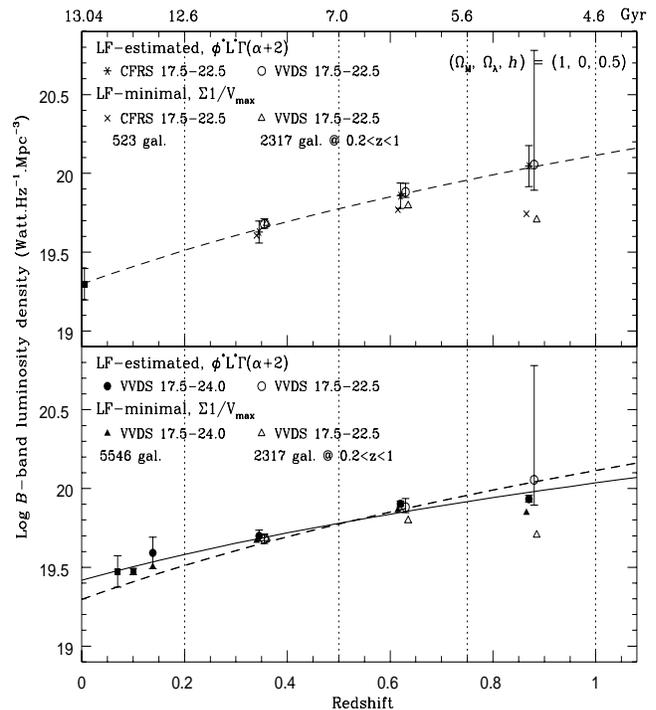}}
\caption{Rest-frame $B$-band comoving non dust-corrected luminosity
  densities with the ($\Omega_{M}$, $\Omega_{\Lambda}$, $h$)~=~(1, 0,
  0.5) cosmology at $0<z<1$ of the $I-$selected CFRS and VVDS surveys.
  Values are listed in Tab.~\ref{table2}. For clarity the $z>0.2$ data
  are represented by symbols slightly displaced horizontally from the
  center of the redshift bins delimited with the vertical dotted
  lines. The plain square points at $z=0.05, 0.07$, and $0.1$ are local
  points from the SAPM \citep{1992ApJ...390..338L}, the 2dFGRS
  \citep{2002MNRAS.336..907N}, and SDSS \citep{blanton03} respectively.
  The dashed line shows the best-fit power law and its associated
  uncertainties for the 'LF-estimated' $\cal{L}_\mathrm{4400~\AA}$ of
  the CFRS 17.5-22.5 data derived in \citet{1996ApJ...460L...1L},
  $(1+z)^{2.72\pm0.5}$, and the solid line is the one for the VVDS
  17.5-24.0 data, $(1+z)^{2.05\pm0.06}$.}
\label{fig6}
\end{figure}

The lower panel of Fig.~\ref{fig6} compares the VVDS
$\cal{L}_\mathrm{4400~\AA}$ cut at $I_{AB}= 22.5$ with the global VVDS
at $I_{AB}= 24.0$ over the same redshift range, $0.2 \le z \le 1.0$,
and with the same cosmology as the CFRS. In the redshift range,
$0.2<z<1.0$, $\cal{L}_\mathrm{4400~\AA}$ from the global VVDS is
consistent within the error bars with the one from the
VVDS-[$17.5-22.5$] (see Tab.~\ref{table2}). Our VVDS local value at
$z=0.138$ is derived by fixing M$^*$ to the SDSS value,
($-19.30-5$log$h$) mag at $z=0.1$ \citep[see Table~2 in][with
$^{0.1}g$-band~$\simeq B$-band at less than a 0.05 mag difference
level]{blanton03}, since we do not span the brightest luminosities at
$z<0.2$. In this redshift range, we observe fainter luminosities than
the SDSS, and we find a steeper faint-end slope than the SDSS
\citep[see discussion in $\S$5.1 in][]{ilbert05}.

The best fit power law for the estimated $\cal{L}_\mathrm{4400~\AA}$
of the VVDS data gives $(1+z)^{2.05\pm0.06}$ to be compared with the
one of the CFRS, $(1+z)^{2.72\pm0.5}$
\citep[see][]{1996ApJ...460L...1L} in cosmology ($\Omega_\mathrm{M}$,
$\Omega_{\Lambda}$, $h$)~=~(1, 0, 0.5). The steeper slope of the CFRS
data is due to the combination of the adopted low local reference at
$z=0$ and the high normalization of the CFRS LF at $z=0.85$ due to a
poor constraint of $\alpha$ and M$^{*}$ with a sample limited at
$I_{AB}=22.5$. 

The CFRS $\cal{L}_\mathrm{2800~\AA}$ evolution was found as
$(1+z)^{3.9\pm0.75}$ \citep[see][]{1996ApJ...460L...1L}. With the same
cosmology ($\Omega_\mathrm{M}$, $\Omega_{\Lambda}$, $h$)~=~(1, 0, 0.5)
and $I_{AB}<24$ the VVDS $\cal{L}_\mathrm{2800~\AA}$ evolves as
$(1+z)^{2.4\pm0.1}$.  The rest-frame $NUV$-2800 is observed in the
optical in the VVDS at $z>0.2$, while in the CFRS it is observed at
$z>0.5$. It adds uncertainties due to the extrapolation of templates.
Thus the $NUV$ CFRS evolution was found too steep as it was already
seen by \citet{wilson02}.

In conclusion we find that the $B$-band luminosity density estimated
from the VVDS is in excellent agreement with that estimated from the
CFRS at $0.2<z<0.75$ (see Tab.~\ref{table2}). We demonstrate that
going deeper in magnitude is superior to assembling larger samples as
far as the galaxy sample is not dependent on cosmic variance and has a
well-defined selection function. Futhermore, comparing the estimates
of the VVDS at 22.5 mag to the ones of the VVDS at 24 mag shows that
the error bars determined in the VVDS are well-defined. Indeed the
estimate at 24 mag is within the error bars of the one at 22.5 mag
since we account for the uncertainty on the LF slope.  Finally, we
confirm that the uncertainties due to template extrapolation led to
the very steep slope of the NUV CFRS luminosity density evolution.
\subsection{Comparison with other deep surveys} 
In Fig.~\ref{fig2} we added the $B$-band luminosity densities from
DEEP2 and COMBO-17 surveys \citep[values taken in Table~2
in][]{faber05}, from the FORS Deep Fields (FDF) survey where Schechter
parameters are taken from Table~A.5 in \citet{gabasch04}, and from the
HDF survey where Schechter parameters are taken from Table~2 in
\citet{poli03}. These surveys at $0.2<z<1$ are in excellent agreement
with the VVDS, except for the \citet{poli03} at $0.4<z<0.7$ which has
a steeper faint-end slope ($\Delta \alpha \sim 0.15 $) than the VVDS,
and thus presents a $B$-band $\cal{L}$ higher by 0.1~dex than the
other surveys in this redshift range. At $1<z<1.2$ the VVDS is higher
by 0.1~dex than DEEP2 because the faint-end slope of the $B$-band LF
is better constrained in the VVDS, a deeper sample by $\sim1$ mag than
DEEP2.  We note that both the $I$-selected VVDS and the deeper by 2.8
mag $I$-selected FDF exhibit a drop of the $B$-band luminosity density
at $z\ga1.1$.

Our 1500~\AA\ $\cal{L}$ results are slightly higher than the ones from
the FDF by 0.1~dex at $0.6<z<2$. It is likely due to their lower fixed
($\alpha=-1.07$) value than our fixed ($\alpha=-1.6$) value. However,
the difference is usually within the error bars.  Our 2800~\AA\
$\cal{L}$ results are in good agreement with the ones from the FDF at
$0.4<z<1.2$, and in excellent agreement with the ones from the SDSS
\citep{baldry05} at $z<0.3$.
\section{Luminosity densities per galaxy type at $z<2$}
\subsection{The VVDS galaxy types}
The large VVDS sample enables us to study the luminosity density
evolution for different galaxy types.  To associate a spectral type to
our galaxies with a known spectroscopic redshift, we used the
best fitting type between UBVRI photometric data and a set of SEDs that 
were lineary interpolated between the four observed spectra of
\citet{coleman80} , i.e. E/S0, Sbc, Scd, and Irr, and two starburst
models from the {\sc GISSEL} library \citep{bruzual93}.  Then we  
divided the galaxy population into four rest-frame color classes, the
elliptical-like ({\it type-1}), the early spiral-like ({\it type-2}),
the late spiral-like ({\it type-3}), and the irregular-like ({\it
  type-4}) types.  We describe in detail in \citet{zucca06} the
fitting process and the robustness of the classification.

We note that several previous deep surveys were limited to the
study of two population sub-samples. For instance in the CFRS,
\citet{1995ApJ...455..108L} studied blue and red populations, simply
dividing the galaxy population into two equal number sub-samples and
corresponding to a Sbc color separation.  This allowed us to identify
little evolution of the red population while the blue population
evolves strongly. In more recent deep surveys, the bimodal rest-frame
color distribution observed at least up to $z=1.5$ is being used to
define an empirical separation between red and blue galaxies (K20;
\citet{fontana04}, DEEP2 and COMBO-17; \citet{faber05}, VVDS;
\citet{franzetti07}). While the bimodality is clearly observed for the
bright galaxies, the faint population does not exhibit two contrasting
modes \citep[see][]{franzetti07}. Also, the bimodality hides a strong
differential, color evolution of the bright population (i.e.
$L>L^{*}$) as shown in Fig.~2 of \citet{zucca06}.  Or, the extensive
analysis of \citet{delapparent03} shows that estimations of LFs based
on two color sub-samples lacks the necessary discriminatory power for
detecting the variations in luminosity as a function of type which are
traced by the intrinsic LFs.  From a theoretical study using
semi-analytical mo\-dels, \citet{kaviraj06} show that the red-sequence
traces the progenitor set of early-type galaxies in terms of numbers
and masses for the bright galaxies, but breaks down severely at faint
($L<L^{*}$) luminosities.  And thus, using the bimodality does not
seem to be robust in classifying galaxies in types unambiguously
related to passively evolving galaxies on one side and star-forming
galaxies on the other.  Whereas, in fact, our selection relies on the
complete SED available from the multi-band imaging. We note that
Seyfert 2 galaxies are included in our sample. We analyse these
galaxies with the true starburst po\-pu\-lation in another paper.
Nevertheless we do not expect a large contribution of the Seyfert 2
population at $z<1$.

\begin{figure}
\resizebox{\hsize}{!}{\includegraphics{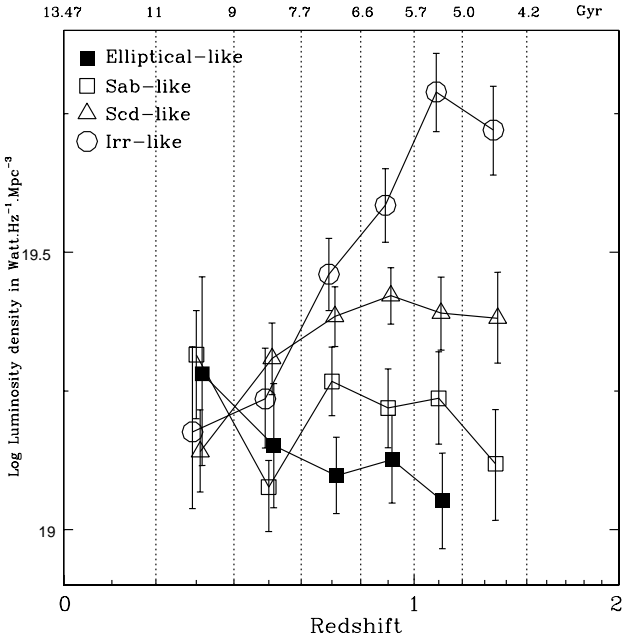}}
\caption{Comoving non dust-corrected luminosity densities in the
  rest-frame $B$-passband from early to late galaxy types (see details
  in Section~6.1). For clarity, data are represented by symbols
  slightly displaced horizontally from the center of the redshift
  bins.  Error bars are at 1$\sigma$. The irregular-like type
  emissivity decreases markedly by a factor 4, while the
  elliptical-like type increases by a factor 1.7 from $z=1.1$ to
  $z=0.2$. We note that at $z<0.4$ the total emissivity is dominated
  by the early-type population.}
\label{fig7}
\end{figure}

We emphasize that our four types are nicely correlated with colors and
asymmetry-concentration parameters from HST images \citep[see Fig.~2
in][]{ilbert06a} and with spectroscopic features \citep[e.g, emission
line strength, 4000~\AA\ break, see Fig.~1 in][]{zucca06}.  The four
individual LFs were derived bet\-ween $0.2<z<1.5$ where our sample is
essentially complete for every type, and Fig.~\ref{fig7} displays the
four luminosity densities in the rest-frame $B$-band where there is
very little extrapolation.
\subsection{Evolution of  $\cal{L}$ per type}
The evolution of the $B$-band LF per type and the evolution of the
fraction of bright ($M_{B_{AB}}-5log(h)< -20$) galaxies has been
described in detail in \citet{zucca06}, and consequently, the
$B$-band luminosity densities evolve as follows.

{\it Type-4: irregular, starburst, very blue galaxies}. The fraction
of bright galaxies decreases from $\sim35$ to $\sim5$ percent from
$z=1.5$ to $z=0.2$. The LF undergoes a strong evolution in density
(negative) and luminosity (negative). There is a strong decrease in
volume density by a factor $\sim2$ coming from both the bright and
faint parts of the LF.  And thus $\cal{L}$ decreases markedly from
$z=1.5$ to $z=0.2$, by a factor $\sim3.5$.

The three other types have an LF which corresponds to a mild evolution
in density (positive) and luminosity (negative).

{\it Type-3: late-spirals, star forming, blue galaxies}.  The fraction
of bright galaxies decreases from $\sim40$ to $\sim10$ percent from
$z=1.5$ to $z=0.2$, and meanwhile the $\cal{L}$ decreases by a factor
$\sim1.7$.  To keep decreasing $\phi^* \times L^*$ this population is
strongly faintening by $\sim1.5$ mag.  $\cal{L}$ decreases markedly at
$z<0.4$ by a factor $\sim1.5$ due to the faintening by $\sim1$ mag of
the LF.

The last two types present an increasing fraction of bright galaxies. 

{\it Type-2: early-spirals, post-starburst, red galaxies}.  The
fraction of bright galaxies increases from $\sim20$ to $\sim35$
percent from $z=1.5$ to $z=0.2$, and globally $\cal{L}$ increases by a
factor $\sim1.6$.  To keep increasing $\phi^* \times L^*$ luminous red
galaxies must appear at low redshifts since this population is
modestly faintening by $\sim0.6$ mag.  Nevertheless there is a
significant variation at low $z$. Indeed, $\cal{L}$ increases from
$z=1.5$ to $z=0.6$, by a factor $\sim1.4$, then decreases from $z=0.6$
to $z=0.4$, by a factor $\sim1.3$, and finally increases from $z=0.4$
to $z=0.2$, by a factor $\sim1.7$. This transition implies that a
small fraction of luminous type-2 galaxies disappears ($\sim$13
percent) while the luminosity decreases by a small factor of
$\sim$0.15 mag.

{\it Type-1: elliptical, red galaxies.} The fraction of bright
galaxies increases from $\sim0.05$ to $\sim55$ percent from $z=1.5$ to
$z=0.2$, and meanwhile $\cal{L}$ increases continuously , by a factor
$\sim1.7$. That is luminous red galaxies must appear at low redshifts
to keep increasing $\phi^* \times L^*$ since this population is
faintening by $\sim$0.3 mag only.

In addition, Fig.~\ref{fig7} shows that the {\it type-1} and {\it
  type-2} red po\-pu\-la\-tions dominate the total light at $z<0.4$, while
at $0.4<z<1.2$ the late-type does. Since we know that red spheroids
are the majority of our {\it type-1} population from our work in
\citet{ilbert06a}, we conclude that a dust deficient population is
dominant at $z<0.4$.  The luminosity density increasing of the {\it
  type-1} and {\it type-2} red populations suggests a contribution
from merging phenomena. Indeed in a downsizing scenario where luminous
red galaxies are already in place at high redshifts ($z\gg1$), and low
luminosity red galaxies appear at low redshifts, a flat luminosity
density would be expected as a function of redshift, whilst adding
merging would increase the luminosity density of the luminous red
population as redshift decreases.
\begin{table*}
  \caption{STY parameters for the rest-frame FUV-1500 LFs of the {\it extended} $2.7<z<5$ data set.}
\label{table3}
\begin{center}
\begin{tabular}{ c c c c c  }
\hline
$<z>^a$ & Redshift range &  $\alpha^b$ fixed & $M^{*}_{AB}(1500)$ mag  & $\phi^{*}$  ($10^{-4}$ Mpc$^{-3}$)\\ 
\hline
            &               &   \multicolumn{3}{c}{Unweighted}             \\
\hline
$3.04$ & [2.70-3.40]  &   $-1.6$ ; $-$1.2 ; $-$1.75 &   $-21.68$ ; $-$21.46 ; $-$21.76 & 6.27 ; 8.02 ; 5.66     \\
$3.60$ & [3.40-3.90]  &   $-1.6$ ; $-$1.2 ; $-$1.75 &   $-22.52$ ; $-$22.38 ; $-$22.84 & 2.64 ; 3.45 ; 1.56     \\ 
$4.26$ & [3.90-5.00]  &   $-1.6$ ; $-$1.2 ; $-$1.75 &   $-22.72$ ; $-$22.44 ; $-$22.86 & 0.63 ; 0.92 ; 0.50     \\ 
\hline
            &               &   \multicolumn{3}{c}{Incompletness-corrected$^c$}             \\
\hline
$3.46$  &  [3.00-4.00] & $-1.4$ ; $-$1.2 ; $-$1.75 &  $-21.38$ ; $-$21.29 ; $-$21.57 & 1.23 ; 1.32 ; 1.00 \\ 
\hline
\end{tabular}
\end{center}
$^a$Mean redshift of the galaxy redshifts in the quoted redshift range. \\ 
$^b$We fixed $\alpha$ to $-1.6, -1.2$, and $-1.75$ and the corresponding STY values for $M^*$ and $\phi^{*}$ are given, respectively. \\
$^c$See \citet{paltani07}. With $\alpha=-1.6$, we have $M^*=-21.49$ and $\phi^*=1.11$.
\end{table*}

We note also that from $z=0.7$ to $z=0.5$, $\cal{L}$ for {\it type-2}
decreases by a factor $\sim1.3$, while $\cal{L}$ for {\it type-1}
increases by a factor $\sim1.15$; this might suggest that $<$15
percent of ellipticals are formed from early-spiral major mergers
within this 1.3 Gyr period.  It might be evidence for mergers between
gas-deficient bright galaxies.  Futhermore, the fact that the {\it
  type-2} population luminosity density increases again from $z=0.5$
to $z=0.3$ by a factor $\sim1.7$ might suggest a density growth due of
an evolution of {\it type-4} galaxies towards {\it type-2} galaxies.

The $\cal{L}$ decreasing of the {\it type-4} population is markedly
different from that of the remaining population. This population is
dominated by dwarf galaxies \citep[see][]{zucca06}.  The evolution of
this population supports a downsizing scenario where most star
formation is shifting to low-mass galaxies at $z<1.2$, while the
global luminosity density is dominated by other galaxy types.
\begin{table}[b]
  \caption{Comoving FUV-1500 luminosity densities at $2.7<z<5$ of the VVDS  for the {\it extended} high-$z$ dataset.}
  \label{table4}
  \begin{center}
    \begin{tabular}{ c c c l  }
      \hline
      $<z>^a$ & Redshift range & $\delta t^b$ & $\log \cal{L}_\mathrm{1500~\AA}$ \\ 
      &                         &                     & W Hz$^{-1}$ Mpc$^{-3}$   \\
      \hline
      &                  \multicolumn{3}{r}{Unweighted minimal}             \\
      \hline  
      $3.04$ & [2.70-3.40]  &   0.6 &   $18.82$          \\
      $3.60$ & [3.40-3.90]  &   0.3 &   $18.95$          \\ 
      $4.26$ & [3.90-5.00]  &   0.3 &   $18.89$          \\ 
      \hline
      &                  \multicolumn{3}{r}{Unweighted estimated}        \\
      \hline  
      $3.04$ & [2.70-3.40]$^c$ & 0.6  &   $19.47^{+0.30}_{-0.33}$          \\
      $3.60$ & [3.40-3.90]$^c$ & 0.3  &   $19.43^{+0.21}_{-0.29}$          \\ 
      $4.26$ & [3.90-5.00]$^c$ & 0.3  &   $18.89^{+0.29}_{-0.29}$          \\ 
      \hline
      &                  \multicolumn{3}{r}{Incompleteness-corrected estimated}        \\
      \hline 
      $3.46$   & [3.00-4.00]$^d$ &  0.5 & $19.47^{+0.37}_{-0.11}$  \\ 
      \hline 
    \end{tabular}
  \end{center}
  $^a$Mean redshift of the galaxy redshifts in the quoted redshift range. \\ 
  $^b$Elapsed time in Gyr in the redshift bin. \\
  $^c$The STY LF estimators are derived using a fixed $\alpha=-1.6$ and error bars are derived from the STY estimations assuming $\alpha=-1.20$ and  $\alpha=-1.75$ as done at $z<2$ (see Tab.~\ref{table1}). \\  
  $^d$In \citet{paltani07}, we use $\alpha=-1.4$. Here, we use the same minimal and maximum values as at $z<2$ for $\alpha$, i.e. $-1.2$ and $-1.75$. With $\alpha=-1.60^{+0.15}_{-0.20}$, we find $\log \cal{L}$~=~$19.64^{+0.20}_{-0.28}$.
\end{table}
\section{Global FUV luminosity densities at $2.7<z<5$}
\subsection{The VVDS high-$z$ population}
In the high-redshift range, $2.7<z<5.0$, the rest-frame 1500~\AA\
corresponds to the observed frames $R$ and $I$ passbands. Thus
uncertainties on absolute magnitudes related to $k$-corrections and
galaxy types are small.  Our data set consists of 161 redshifts with
{\sc VVDS} quality flags 2, 3, and 4, and 237 redshifts with {\sc VVDS}
quality flags 1 in the VVDS-0226-04 field ($\sim$0.5~deg$^2$).  The
latter fraction is not negligible since the difficulty in determining a
redshift is increased at the faintest apparent magnitudes.  The
confidence levels for the single high-$z$ population are $>50$ percent
for the flags 2, 3, and 4 dataset, and $\sim45$ percent for the flags
1, 2, 3, and 4 dataset \citep{lefevre05b}.

Here, we correct our sample for the target sampling rate only (see
Section~2).  That is, there is no assumption for the redshift
distribution of the sources that were not spectroscopically observed
or for which no redshift could be reliably identified from the
spectrum obtained. This is a more restricted approach than what was 
adopted in \citet{ilbert05}, where the photometric redshifts were used
to obtain a spectroscopic success rate as a function of redshift, and
thus to correct further for incompleteness.  We could not use the
same approach here because our photometric redshifts have been thoroughly
tested only up to $z\sim2$ \citep{ilbert06c}.  The correction for
incompleteness has for effect to steepen the slope of the unweighted
LF, and since $\alpha$ is correlated with $M^*$, it produces a
brightening \citep[see Fig.~4 in][]{ilbert05}.

We consider the two following high-$z$ datasets.  The {\it standard}
dataset is composed of quality flags 2, 3, and 4 redshifts and the
{\it extended} dataset is composed of quality flags 1, 2, 3, and 4
redshifts. It is likely that the true luminosity function/density lies
between these two cases, assuming that the 7 percent of flags 0 over
the whole $0<z<5$ sample makes little difference.  Using one field of
view is likely to induce uncertainty due to cosmic variance.
Nevertheless \citet{sawicki06a} observed an effective area of
169~arcmin$^2$, i.e. one third of ours, 410 arcmin$^2$, split into
five fields (the Keck Deep Fields; KDF) and found a quite moderate
cosmic-variance effect.

In \citet{paltani07} we do a comprehensive study about the impact of
the large uncertain redshift population at $3<z<4$, in using a
somewhat different analysis. Here we choose to use the same analysis
as at $z<2$. We note that results are qualitatively similar.
\subsection{Estimation of the high-$z$ FUV VVDS luminosity functions and densities}
\begin{figure}[t]
\resizebox{\hsize}{!}{\includegraphics{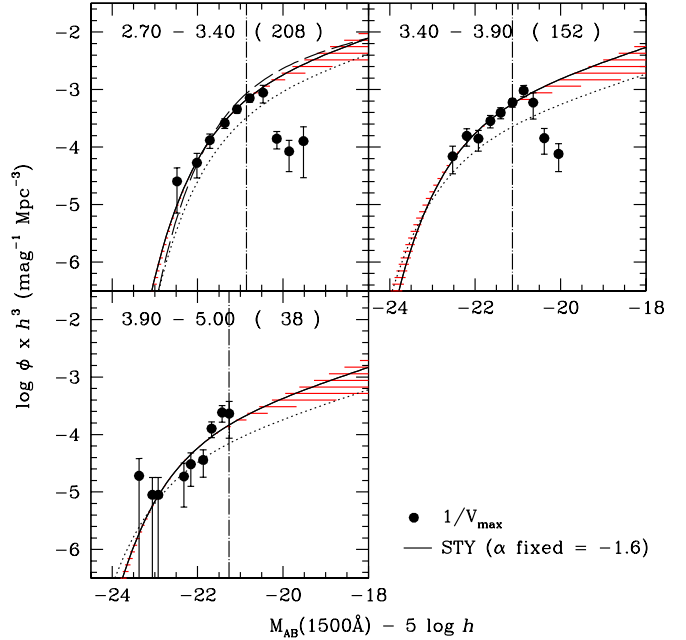}}
  \caption{Unweighted non dust-corrected luminosity functions in the
    rest-frame FUV-1500 passband estimated with $\alpha$ fixed to
    $-1.60$.  The circles and the solid line represent the V$_{max}$
    and the STY estimates for the {\it extended} dataset where the
    number of galaxies is given within parenthesis next to the
    redshift range.  The vertical dot-dashed line corresponds to the
    faint absolute magnitude limits considered in the STY estimate
    (see $\S$~3.2). The hashed area is limited by the LFs derived with
    $\alpha$ fixed to $-1.20$ and $-1.75$. The dotted lines represent
    the STY estimates with $\alpha$ fixed to $-1.60$ for the {\it
      standard} dataset. The {\it standard} dataset LF is below the
    {\it extended} dataset LF since it does not include quality flags
    1 (see $\S$7.1). In the top-left panel, the long-dashed line is
    the incompleteness-corrected dataset FUV-1500 LF at $3<z<4$
    derived as in \citet{paltani07}, with fixed $\alpha=-1.4$.}
  \label{fig8}
\end{figure}

\begin{figure}
\resizebox{\hsize}{!}{\includegraphics{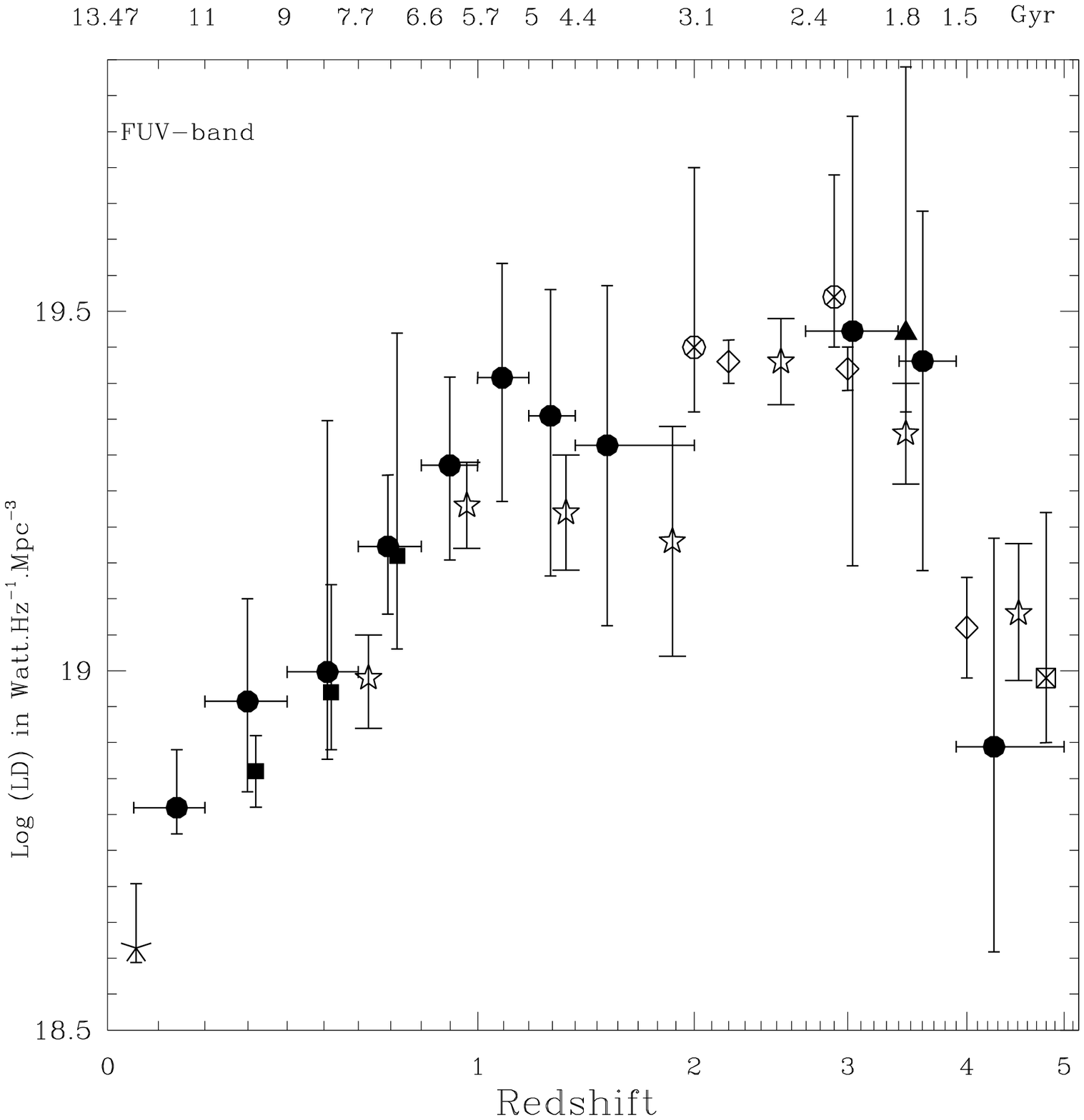}}
\caption{Comoving non dust-corrected FUV luminosity densities from
  $z=0$ to $z=5$. The {\it plain circles} represent the 1500~\AA\ VVDS
  data ($0.05<z<5$) (see values and error bars in Tabs.~\ref{table1}
  \&~\ref{table4}), the {\it plain triangle} represents the 1500~\AA\
  VVDS data at $3<z<4$ derived as in \cite{paltani07}, and the {\it
    plain squares} represent the 1500~\AA\ GALEX-VVDS data at $z<1$,
  plotted at $+0.02$ in redshift for clarity, from
  \citet{schiminovich05}.  Other data are the 1500~\AA\ GALEX-2dFGRS
  from \citet{wyder05} (asterisk), the 1500~\AA\ HDF data from
  \citet{arnouts05} (crossed circles), the 1700~\AA\ KDF data from
  \citet{sawicki06b} (open rhombus), the 1500~\AA\ FDF data from
  \citet{gabasch04} (open stars), and the 1700~\AA\ data from
  \cite{iwata07} (crossed square).  }
  \label{fig9}
\end{figure}

For the rest-frame FUV analysis, we fixed $\alpha$ to $-1.6$.  As
shown in Fig.~\ref{fig1} the uncertainty in a slope $\alpha <-1.3$ may
lead to large discrepancies in the estimation of $\cal{L}$. For
consistency, we have derived error bars of the rest-frame FUV
$\cal{L}$ from the STY estimations assuming $\alpha = -1.20$ and
$\alpha = -1.75$ for the {\it extended} high-$z$ data set as we did at
$z<2$.  The LF parameters are given in Tab.~\ref{table3}.
Fig.~\ref{fig8} displays the LF estimations of the {\it standard} and
{\it extended} high-$z$ datasets, and the incompleteness-corrected LF
estimated at $3<z<4$ in \citet{paltani07}.  We do a detailed
comparison for different fixed $\alpha$ values found in the literature
in \citet{paltani07}.  The rest-frame FUV $\cal{L}$ at $2.7<z<5$ are
displayed in Fig.~\ref{fig9} together with those at $z<2$, and they
are listed in Tab.~\ref{table4}.
\subsection{Comparison with other high-$z$ surveys}
Fig.~\ref{fig9} displays other rest-frame FUV luminosity densities
from the literature.  At $2<z<5$, our $I$-selected rest-frame FUV
$\cal{L}$ are broadly in agreement with the other estimations. At
$z\sim3$, rest-frame FUV $\cal{L}$ estimations are found within
0.1~dex (19.47 for the VVDS, 19.58 for \citet{steidel99}, 19.43 for
\citet{gabasch04}, 19.52 \citet{arnouts05}, and 19.51 for
\citet{sawicki06b}).  We note that the lumino\-si\-ty functions exhibit
noticeable discrepancies, in particular at the bright end, and
furthermore that the faint end slope is not constrained \citep[see
detailed comparisons in][]{paltani07}. As at $z<2$, our high-$z$
luminosity densities are slightly higher than those estimated from the
FDF (due to their lower adopted slope of $\alpha=-1.07$) except at
$3.9<z<5$.  The decrease from $z=3.4$ to $z=5$ is smaller in the FDF
(0.25~dex) than in the VVDS (0.5~dex), however our error bars are
compatible with the FDF result. This may suggest that our unsuccessful
spectroscopic identification rate has a strong effect in our highest
VVDS redshift bin, while our data points at $2.7<z<3.9$ appear fully
consistent with the FDF.
\section{The FUV luminosity densities at $0<z<5$}
\subsection{The global shape} 
Fig.~\ref{fig9} displays the rest-frame FUV luminosity densities
within the redshift range $0<z<5$.  We find that the global luminosity
density increases by a factor $\sim3.5$ from $z=0.05$ to $z=1.2$, by a
factor $\sim1.2$ from $z=1.2$ to $z=3.4$, and decreases by a factor
$\sim0.3$ from $z=3.4$ to $z=5$. Futhermore the evolution at
$1.1<z<3.0$ might be more complex than a modest increase. Indeed, even
though our large error bars cannot exclude an increase or a plateau,
our data points taken at face value exhibit a decrease by a factor
$\sim1.3$ from $z=1$ to $z=2$, and an increase by a factor $\sim1.4$
from $z=2.0$ to $z=3.4$.  One could mention dust attenuation effects
since at these redshifts, targets are selected from their rest-frame
UV, and thus a fixed $I$-band flux cut could miss a non-negligible
fraction of highly dust enshrouded targets. However, UV luminosity
densities at $z\sim3$ are larger than at $1<z<2$ while they should be
even more dust affected since selected at shorter ultraviolet
wavelengths.

The 1500~\AA\ luminosity density as a function of redshift from the
FDF \citep{gabasch04} presents a similar shape to that derived from
the VVDS. We stress that both datasets, the FDF and the VVDS, span 
the redshift range $0.5<z<5$ within one single survey and have the
same unique $I$-band selection criterion.  We note that the FDF used
photometric redshift techniques with NIR photometry, which gives
reliable redshifts at $1<z<2$.  The VVDS and the FDF are
complementary; the VVDS is $\sim50$ times larger in surface than the
FDF and it consists of spectroscopic redshifts, while the FDF goes 2.8
mag deeper than the VVDS using photometric redshfits. Both estimates
are in agreement, in particular at $1<z<2$.

We note that the empirical models from \citet{perez05} also show a
decrease in the cosmic SFR density bet\-ween $z=1.4 $ and $z=2.2$,
varying from 10 to 29 percent depending on the model but with
luminosity evolution solely. It is interesting to see that their model
including a combined luminosity plus number density evolution does not
exhibit a decrease. Rather than excluding a number density evolution
it indicates that the global emissivity is dominated by the luminosity
evolution. We go into more detail about this effect in the next
section.

We emphasize that our $I$-selected sample misses the very red
$(I-K)>2.6$ galaxy population at $1.2<z<2.5$ . This po\-pu\-lation
concerns the most massive star forming dusty galaxies, as detailed in
our $K$-selected sample described in \citet{pozzetti07}.
\subsection{The intrinsic luminosity dependency} 
\begin{figure}[t]
\resizebox{\hsize}{!}{\includegraphics{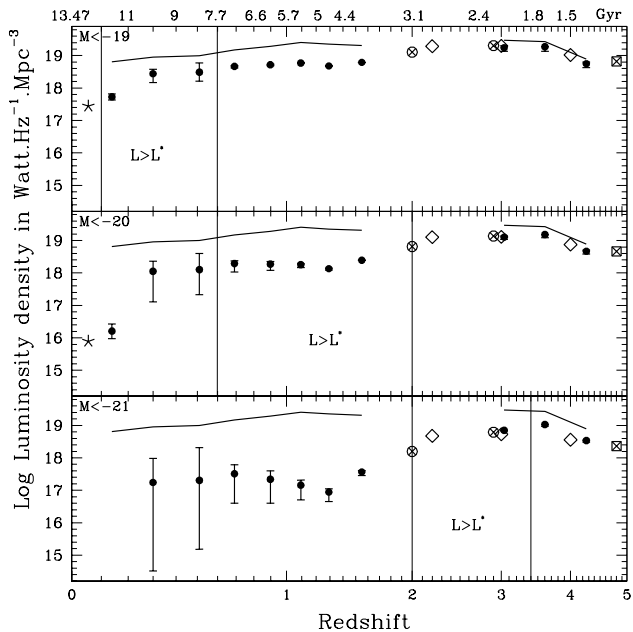}}
\caption{Comoving rest-frame FUV non dust-corrected luminosity
  densities from $z=0$ to $z=5$ for three bright populations defined
  as $M_{AB}(1500~\AA)<-19, -20$, and $-21$ mag. Symbols are the same
  as in Fig.~\ref{fig9}. Error bars correspond to the fixed slopes,
  $\alpha=-1.2$ and $-1.75$ at $0.2<z<5$, at $0.05<z<0.2$ to the fixed
  M$^{*}\pm0.1$; those which are not visible are smaller than the size
  of the data point, i.e. $<0.1$~dex. We do not plot the local values
  for $M<-21$ which have low values, at $z=0.14$ and $0.055$
  Log~$\cal{L}$~=~$12.45\pm0.55$ and 12.10 Watt~Hz$^{-1}$~Mpc$^{-3}$,
  respectively. In each panel, the solid line connects the VVDS points
  of the global FUV luminosity density as displayed in
  Fig.~\ref{fig9}, and the vertical lines delimit the approximative
  redshift range in which the bright population is more luminous than
  the characteristic FUV luminosity of the whole population.}
\label{fig10}
\end{figure}

\begin{figure}
\resizebox{\hsize}{!}{\includegraphics{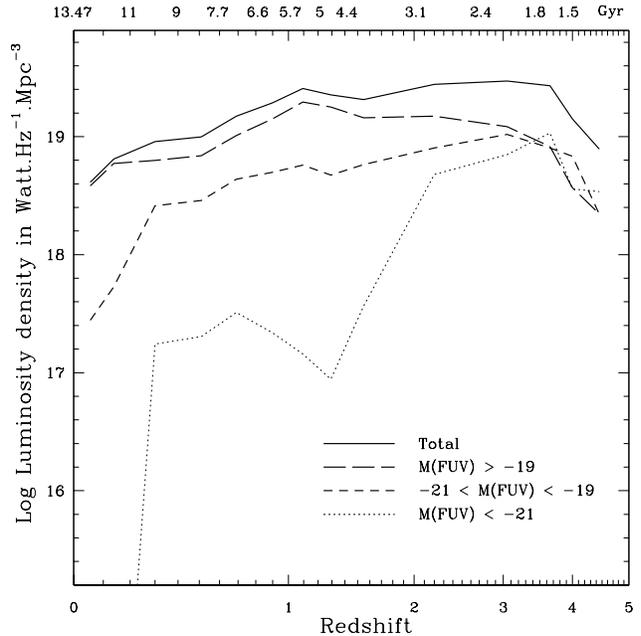}}
\caption{Comoving rest-frame FUV non dust-corrected VVDS luminosity
  densities from $z=0$ to $z=5$ for three luminosity-class populations
  defined as $M_{AB}(1500~\AA) > -19$ (dwarf and
  intermediate-luminosity galaxies, $< 0.1\ {\rm L}^*_{\rm LBG}$),
  $-21 < M_{AB}(1500~\AA)< -19$ (luminous galaxies), and
  $M_{AB}(1500~\AA) < -21$ (very luminous galaxies, $> {\rm L}^*_{\rm
    LBG}$ ). We used and connected the same VVDS dataset as displayed in
  Fig.~\ref{fig10}.}
\label{fig11}
\end{figure}

The luminosity density is dominated by the evolution of the bright
population.  Thus we have integrated the LF from the three following
absolute magnitude limits: $M_{AB}(1500~\AA)<-19$ mag,
$M_{AB}(1500~\AA)<-20$ mag, and $M_{AB}(1500~\AA)<-21$ mag.  According
to our LFs, these magnitude cutoffs correspond about to $L>L^{*}$
galaxies at $0.1<z<0.6$, $0.6<z<2$, and $2<z<3.4$ respectively. At
$z\sim0.05$, $M_{AB}^{*}(1500~\AA)\sim-18$ mag \citep{wyder05}.  We
choose a fixed cutoff rather than a luminosity evolving cutoff to make
a comparison over a long time baseline because this does not depend on
the mo\-de\-ling of the luminosity evolution of the LF.  The change of
M$^*$ with redshift implies that at $z\simeq3$ the most luminous
galaxies were forming stars at a rate $\sim$5, and $\sim$7 times
higher than at $z\simeq$~0.15, and $z\simeq$~0.05 respectively.  We
have applied the same cutoffs to the LF integration of the 1500~\AA\
HDF \citep{arnouts05}, the 1700~\AA\ KDF \citep{sawicki06b}, and the
1700~\AA\ data point of \citet{steidel99}. Fig.~\ref{fig10} displays
these different luminosity densities for the bright population.  We
also produced the rest-frame FUV-band luminosity densities for three
range of luminosities, $M_{AB}(1500~\AA)>-19$ mag (dwarf and
intermediate population, $< 0.1\ {\rm L}^*_{\rm Lyman\ Break\
  Galaxy}$), $-21<M_{AB}(1500~\AA)< -19$ (luminous population), and
$M_{AB}(1500~\AA)<-21$ (very luminous population, $> {\rm L}^*_{\rm
  LBG}$), as shown in Fig.~\ref{fig11}.
\subsubsection{Detailed results} 
We observe that on average the rest-frame FUV-band lumi\-no\-si\-ty density
of the ($2.7<z<5$) population is significantly brighter by a factor at
least $\sim6$ than the ($0.2<z<1.4$) population. This implies a
transition phase for the population dominating the FUV emissivity.
Another transition phase is observed at $z<0.2$ with a steep decline
by a factor of at least $\sim10$ of the bright population luminosity
density.  Concentrating on the most luminous galaxies, and taking the
values of $\cal{L}$ at their face value, we can tentatively identify
five phases from $z=5$ to $z=0$ as follows.

{\it (a) From $z=5$ to $3.4$}, the emissivity due to galaxies brighter
than $M_{AB}(1500~\AA)<-20$ and $-21$ mag increases by a factor of $\sim3$ and 
2.5 respectively.  The whole population, from the dwarf to the very
luminous galaxies, sees its FUV luminosity increasing. This
corresponds to a very active phase in terms of newly formed stars in
every galaxy.

{\it (b) From $z=3.4$ to $1.4$}, the emissivity due to galaxies
brighter than $M_{AB}(1500~\AA)<-20$ and $-21$ mag steadily decreases by a
factor $\sim6$ and 25 respectively. This $\sim3$Gyr phase corresponds to
the progressive, but relatively quick, drop of star formation in the
most luminous galaxies which are formed earlier than $z=3.4$, and
whose contribution to the global FUV luminosity density becomes less
important toward lower redshifts.  \cite{lotz06} observe that the
distribution of the HST rest-frame FUV morphologies at $z\sim1.5$ are
similar to the ones at $z\sim4$ with an identical fraction of
major-merger candidates.  This suggests that this drop is due to the
same luminous galaxy population, and that it is not caused by a
decrease in the number of mergers.  Thus from $z=3.4$, the most FUV
luminous galaxy population has suddently finished its stellar mass
assembly and active star forming phase, while the less FUV luminous
population is still very active and becomes the dominant star forming
galaxy population until nowdays.

{\it (c) From $z=1.4$ to $0.6$}, the emissivity due to galaxies
brighter than $M_{AB}(1500~\AA)<-20$ and $-21$ mag increases by a factor
$\sim2$ and 4 respectively.  This corresponds to Fig.~2 in
\citet{zucca06}, where the fraction of the bright population (i.e.
$L>L^{*}$) of early-type (E- and Sab-like) galaxies increases while
the one for late-type (Scd-, Irr-like) galaxies decreases even though
the latter still dominate in terms of numbers and also in emissivity
(see Fig.~\ref{fig7}) up to $z\sim0.7$. This phase could correspond to
merger events; indeed very disrupted HST morphologies in the very
luminous galaxy population are generally observed.  In the CDFS, we
observe mergers within the early-type population as shown in Fig.~2 of
\citet{ilbert06a} and also that the volume density of red, bright
bulge-dominated galaxies increases by a factor $\sim2.7$ from $z\sim1$
to $z\sim0.7$.  Furthermore from $z=1.5$ to $z=0.6$ we find in
\citet{ilbert06b} that rest-frame $B$-band $\cal{L}$ for galaxies
lying in over-dense environments (at a scale of 5$h^{-1}$ Mpc)
increases by a factor $1.2$, while the one for those in under-dense
environments continuously decreases by a factor of 2.3. Since the most
luminous galaxies are usually found in dense environments, our results
favor the merging event to build-up early-type galaxies.  The exact
percentage of merger events is still hotly debated. We note also that
it corresponds also to the phase where the less luminous population
has reached its peak of star formation activity at $z\sim1.1$.
Another point is that at $1.5<z<2.5$, our $I$-selection is missing the
reddest galaxies \citep{pozzetti07}, the dusty and massive sources as
discovered in infrared surveys (i.e., e.g. \citet{daddi04}), and
undergoing strong star forming and dusty phase. This population would
be an excellent contributor to populate the FUV luminous population at
$z<1.4$ once the dusty phase linked to intense burst does not dominate
anymore.

{\it (d) From $z=0.6$ to $0.2$}, the emissivity due to galaxies
brighter than $M_{AB}(1500~\AA)<-20$ and $-21$ mag decreases by a factor
$\sim1.3$ and 1.5 respectively.  The decrease of the bright part of the
emissivity corresponds to a phase entirely dominated by the early-type
galaxies for which the star formation decreases passively.

{\it (e) From $z=0.2$ to $0.05$}, the emissivity due to galaxies
brighter than $M_{AB}(1500~\AA)<-20$ and -21 mag decreases by a factor
$\sim100$ and 25 respectively; the bright part of the global FUV
luminosity density abruptly drops.  Analysing the $0<z<0.3$ CFRS
spectra, \citet{tresse96} found that only 54 percent of galaxies
exhibit strong star formation with H$\alpha$ and H$\beta$ both in
emission and 15 percent have H$\alpha$ and H$\beta$ both in
absorption. The remaining 31 percent is the intermediate population
which shows spectral features from both star burst and quiescent
stellar evolution.  Therefore star formation seems to have slowed down
or stopped in a significant fraction of galaxies in the nearby
Universe.  This population might correspond to the descendants of
$L^*$ galaxies at $0.2<z<1.4$ which when they fade in luminosity, stop
dominating the rest-frame FUV-band luminosiy densities. We note that
the local sudden drop might be combined with the local underdensity
seen in optical redshift surveys (\citet{zucca97}, \citet{tresse98}).

In resume, we observe a first major crisis for producing efficiently
new stars, lasting $\sim3$ Gyr, from $z\sim4$ to $z\sim1$, which
involves the most massive and luminous galaxies
($M_{AB}(1500~\AA)<-21$, $M_{*}>10^{11} M_{\sun}$).  While gradual
fading of the global population starts at $z\sim1$, we observe a
second major crisis which started $\sim3$ Gyr ago when star formation
was progressively stoping in intermediate mass and luminosity galaxies
($-20<M_{AB}(1500~\AA)<-19$, $10^{9}<M_{*}<10^{11} M_{\sun}$).  The
period from $z=1.4$ to $z=0.6$ is highly intricate due to
intertwining populations that follow different evolutionary path.
Nevertheless the combined effects of decreasing both luminous and
intermediate population is to make the global luminosity density
decrease faster at $z<1.2$ than at $1.2<z<3.9$.
\subsubsection{Evolution of $L>L^{*}$ galaxies} 
According to the Schmidt law of star-forming galaxies
\citep{kennicutt98}, the SFR is scaled to the cold gas density to the
1.4 power. Thus, once the gas reservoir is exhausted, star formation
ceases. The old, most luminous and massive galaxies have exhausted
their gas reservoir during their early intense star formation
$z\ge3.5$ (see previous $\S$), and since then they undergo passive
evolution as star formation ceases.  This creates excellent dry
candidates, that is cold gas-depleted to prevent new star formation.

Creation of new galaxies occurs as the threshold amplitude for forming
bright galaxies decreases as described in \citet{marinoni05} from the
VVDS data. That is, the typical $L^*$ of the population created at a
given $z$ will decrease with decreasing redshift. This implies that
younger, less massive and less luminous $L^{*}$ galaxies continue to
efficiently form stars with a large reservoir of cold gas at
$0.2<z<3.9$. This intermediate galaxy population presents a peak of
SFR at $z\sim1$.  And at $z<0.2$ these galaxies appear to have also
exhausted their gas as suggested by the 1500~\AA\ $\cal{L}$ drop, and
start to evolve passively. The gas-exhaustion would favor the
evolution of morphologies toward early-type galaxies.

In resume, the global FUV luminosity density decreases faster at
$z<1.2$ than at $z<3.9$ since it combines the decrease from both
luminous and intermediate populations. During the phase $1.2<z<3.9$ it
undergoes a decline of 0.06~dex, while for the phase $0.2<z<1.2$ it is
0.45~dex. The small decrease corresponds to the phase where the dwarf
and intermediate galaxy population increases its SFR, while the
luminous population start to undergo its SFR decline. In
Fig.~\ref{fig9}, the global luminosity density might decrease by
$\sim0.1$~dex from $z=3.9$ to $z\sim2$, then increase by $\sim1.2$~dex
from $z\sim2$ to $z=1.2$.  Nevertheless, our large error bars do not 
exclude a smooth decrease by $0.06$~dex from $z=3.9$ to $z=1.2$.

At $z\la 4$, both processes, i.e. dry mergers toward decreasing
redshift and morphologies evolving toward early-type galaxies, might
contribute to an increase of the bright early-type population by a
factor $\sim10$ to reach $\sim55$ percent of the total population from
$z=1.5$ to $z=0.2$, while the early-type population undergoes a
passive luminosity evolution as shown by \citet{zucca06}.  It is in
agreement with semi-analytical studies, e.g. \citet{kaviraj06} find
that less than 50 percent of the stellar mass which ends up in
early-types today is actually in early-type progenitors at $z\sim1$,
or \citet{delucia06} find that 50 percent of local elliptical accrete
half of their stellar mass out $z\sim0.8$.
\section{The history of the star formation rate density} 
\begin{figure}[t]
\resizebox{\hsize}{!}{\includegraphics{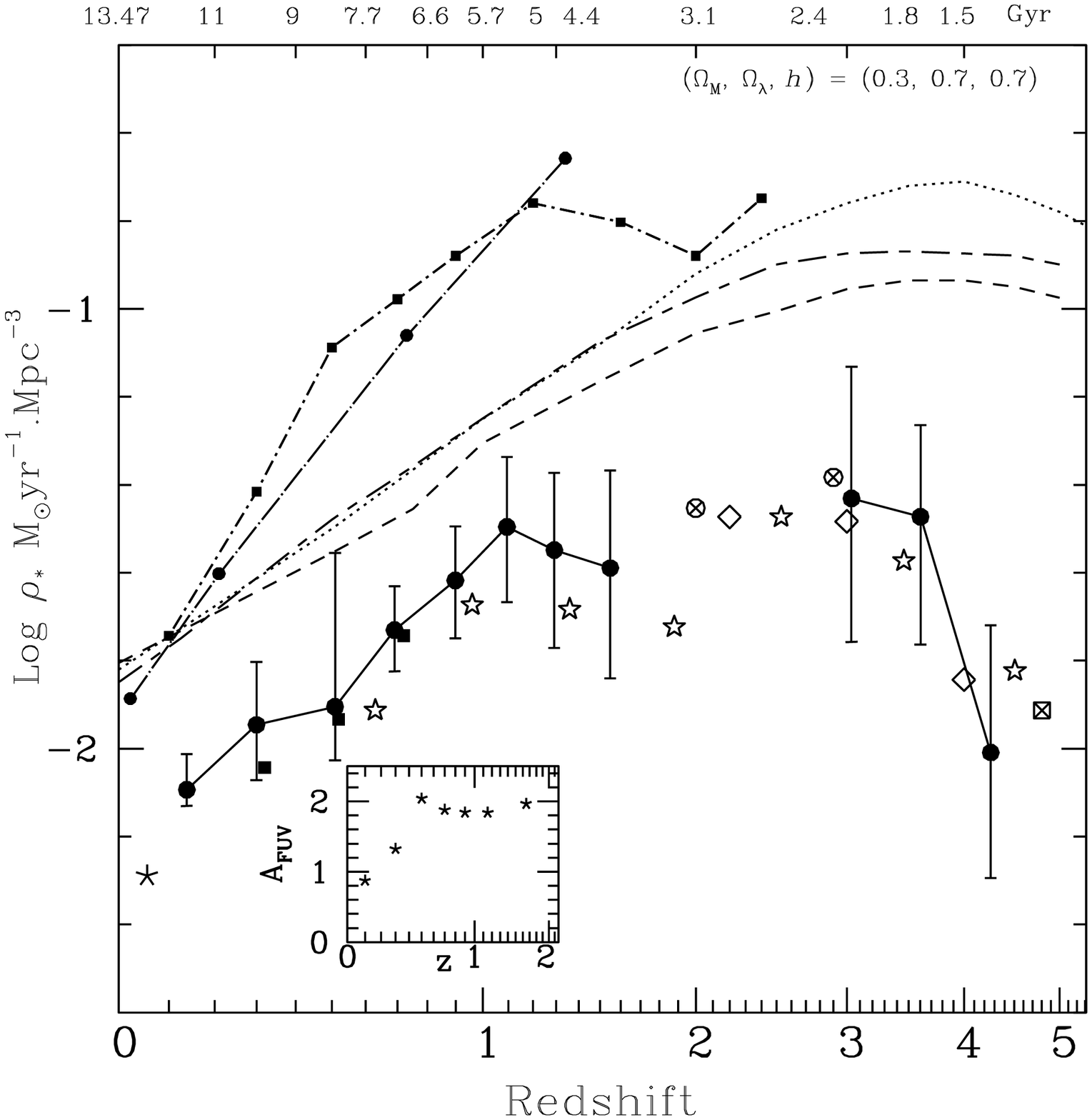}}
\caption{Star formation rate densities as a function of redshift. Symbols ({\it plain large circles, plain large squares, asterisk, crossed circles, open rhombus, open stars, crossed square}) and error bars are the same as in Fig.~\ref{fig9}, and they represent the SFR densities with no reddening correction.  The dotted, long-dashed line connects the reddening corrected H$\alpha$ data (small plain circles) from the CFRS \citep{tresse98,tresse02}. The dotted, short-dashed line connects the 12$\mu$m data (small plain squares) from \citet{perez05}.  Models from \citet{somerville01}, \citet{croton05} and \citet{nagamine06} are represented with the dashed, the dotted, and the short-long dashed lines respectively. Everything is for a \citet{salpeter55} IMF. The derived $A_{\rm FUV}$ dust obscuration between the VVDS data not dust corrected and the 12$\mu$m data at $z<2$ is represented in the inset.}
\label{fig12}
\end{figure}

We derive the SFR densities hich are not dust corrected using the rest-frame
1500~\AA\ luminosity densities of the VVDS from $z=0$ to $z=5$.  The
SFR calibration of \citet{madau98} yields $SFR = 8 \times 10^{27}
L(1500~\AA/\rm erg~s^{-1}~Hz^{-1})~M_{\bigodot}~yr^{-1}$ for a
\citet{salpeter55}'s IMF including stars from 0.1 to 125 solar mass.
We recall that at $0<z<1$, our 1500~\AA\ rest-frame band spans
ultraviolet wavelengths shorter than 3000~\AA, that is a non-observed
wavelength range with the optical bands.  Nevertheless, we 
checked that our results are fully consistent with the rest-frame
1500~\AA\ GALEX-VVDS data (see details in Section~4.2). We present our
SFR density in Fig.~\ref{fig12}.
\subsection{Dust obscuration} 
Our FUV data are not dust corrected. In the following, we compare them
to data which account for dust at $z<2.0$ to estimate the global
correction needed to recover the total SFR density.  That is, we use
the H$\alpha$ nebular emission line dust-corrected data taken from \citet{tresse02},
and the 12$\mu$m mid-infrared continuum data from \citet{perez05}.
These results are both described with an evolution proportional to
$\sim(1+z)^{4}$, while the evolution of 1500~\AA\ $\cal{L}$ in the
VVDS is best described by $\sim(1+z)^{2}$ (see Section~4.1).  Assuming
an average dust correction for the whole field galaxy population, the
dust obscuration at 1500~\AA\ is $\sim1.8-2$ mag from $z=2$ to
$z=0.4$.  And from $z=0.4$ to $z=0$, it becomes smaller from $\sim2$
down to $\sim0.9-1$ mag.

The apparent dust transition at $z<0.5$ corresponds exac\-tly to the
change of the dominant population as seen in Fig.~\ref{fig7} where the
early-type galaxy $B$-band emissivity starts to do\-mi\-nate below
$z=0.4$. The early-type population is known to be dust deficient, and
thus the needed amount of dust obscuration is less strong as redshift
decreases. Our argument assumes that the increasing early-type
dominant population in the $B$-band emissivity holds in the $UV$-band
emissivity.  From $z=0.5$ to $z=2$, the $B$-band emissivity is
dominated by the late-type population, which form stars from gaseous
nebulae, and thus is attenuated by a constant factor. From $z=2$ to
$z=4$, the star-forming high-mass galaxies still dominate the
emissivity, and thus we expect a constant dust attenuation of $2$ mag.
We note that near-infrared surveys miss the faint blue galaxies at
$z<1.2$, while they include the extremely red galaxies at $z>1.2$ as
shown in \citet{pozzetti07}.  Still the 12$\mu$m data from
\citet{perez05} exhibit a similar shape than the FUV data, which
reinforces the no evolution of the dust content as shown by
\citet{bell04} at $z<1$ and \citet{reddy06} at $z>1$.  The dust
correction might change again at $z>4$ where the by-products of the
physical processes to actively form star in massive galaxies will
dominate through AGN and SN feedback.

We will make detailed dust obscuration estimates at $0<z<2$ in future
papers using the VVDS-GALEX and -SWIRE data.
\subsection{Comparison to simulations} 
In Fig.~\ref{fig12} we display the simulations from
\citet{somerville01}, \citet{croton05}, and \citet{nagamine06}.
Comparing simulations with observational measurements might be
severely affected by the assumed IMF, and the dust obscuration.  These
three simulations use the stellar population synthesis models of
\citet{bruzual93}. \citet{nagamine06} uses the IMF of
\citet{chabrier03}, while the others use the one of
\citet{salpeter55}. The SFR calibration of \citet{nagamine06} yields a
conversion factor of $1.24 \times 10^{28}$ for a \citet{chabrier03}
IMF including stars from 0.01 to 100 solar mass. It differs by
$\sim$0.2 in log from a \citet{salpeter55} IMF, and thus, for
consistency, we add this factor to the simulations of
\citet{nagamine06}.  The dust prescription differs in a complex manner
in each simulation. We do not attempt to homogenize the simulations in
terms of dust obscuration.

The simulations all exhibit a peak of the SFR density at $z\sim4$ with
a smooth decrease up to $z=0$. At a quick glance, they do not seem to
go through the observational points, either dust corrected or not.
However, at $2<z<3.9$, the simulations are close to our FUV data taken
at face value if we assume a constant dust attenuation of $\sim2$ mag
as discussed in the previous paragraph.  And, at $0<z<0.2$, the
simulations are in good agreement with the dust corrected data
(H$\alpha$ and 12 $\mu$m), or with FUV dust-corrected data by 1 mag.
We note that \citet{nagamine06} used the dust extinction factors
of \citet{steidel99}, i.e. 1 mag for $z<2$ and 1.7 mag for $z>2$.
Still, at $0.2<z<2$, the simulations do not reproduce the
observational data.  At $z<0.2$ and $z>2$, the global FUV emissivity
is dominated by luminous, massive, large galaxies, while at $0.2<z<2$
it is dominated by the intermediate population.
\section{Conclusions}
We studied the first epoch VVDS data purely $I$-selected at
[17.5-24]~mag in (AB).  The sample is unique in the sense that it goes
deeper than previous $I$-selected spectroscopic samples, it has a
well-defined selection function and it has enough data to study
sub-samples. Within a single survey, we trace the evolution of the
galaxy population dominating the total light at different redshifts
all the way from $z=5$ to $z=0$.  The main results of our
comprehensive study are summarised below.
\begin{itemize}
\item[i)]To study the luminosity density evolution, observing deeper
  in flux is superior to assembling large samples as far as the galaxy
  sample is not dependent on cosmic variance and has a well-defined
  selection function.  Nevertheless, the survey field must be large
  enough to identify the rare bright objects required to constrain the
  bright end of the LF.
\item[ii)]The luminosity density evolution is substantially
  wavelength-dependent since the rest-frame passband luminosity is
  related more or less directly to very different stellar populations.
  We find that the non dust-corrected $\cal{L}$ evolve with time over
  $0.05\le z \le1.2$, as $\cal{L} \propto \rm (1+z)^{x}$ with $x=2.05,
  1.94, 1.92, 1.14, 0.73, 0.42$, and $0.30$ in the FUV-1500, NUV-2800,
  U-3600, B-4400, V-5500, R-6500, and I-7900 passbands, respectively.
  Although error bars are still large, we note that most luminosity
  densities exhibit a transition at $z\simeq1.1$ in the evolutionnary
  tendency. Over the last 8.5~Gyrs, the SFR-related FUV-band
  luminosity density drops by a factor 4. Recent merger events should
  produce little star formation.
\item[iii)]The $B$-band luminosity density evolution is strongly
  type-dependent. From $z=1.2$ to $z=0.2$ the irregular-like, and
  Scd-like type galaxies decrease markedly by factors 3.5, and 1.7,
  respectively while the elliptical-like, and Sab-like types increase
  by factors 1.7, and 1.6, respectively. The late-type galaxy
  population undergoes a downsizing scenario where most star formation
  is shifting to intermediate- and low-mass galaxies at $z<1.2$.  The
  early-type galaxy population suggests a little contribution from
  merging or simple rest-frame color evolution phenomena, but it
  evoles passively in terms of luminosity, which suggests a very early
  formation stage ($z\gg4$) for the majority of this red population.
<\item[iv)]The SFR density as seen in the rest-frame FUV-1500
  luminosity density without dust correction undergoes the several
  following up-and-down phases as redshift decreases, and is strongly
  luminosity dependent.  From $z=5$ to $z=3.4$, it increases by at
  most a factor $\sim3.5$, and it corresponds to the end of the mass
  assembly of the most luminous galaxies, $M_{AB}(1500~\AA)<-21$
  (L$>$L$^*_{\rm LBG}$), which present a peak of SFR at $z\simeq3.5$.
  From $z=3.4$ to $z=1.2$ it decreases by a factor $1.2$.
  Notwithstanding the large error bars this small decrease might be
  the result of two opposite processes; a decrease by a factor
  $\sim1.4$ from $z=3.4$ to $z=2$ due to the fading of the giant
  galaxy population by a factor $40$, and an increase by a factor
  $\sim1.3$ from $z=1.2$ to $z=1.0$ due to the shift of the star
  formation activity towards the less luminous
  ($-21<M_{AB}(1500~\AA)<-19$) galaxy population which presents a peak
  of SFR at $z\sim1$.  From $z=1.2$ to $z=0.05$ it declines steadily
  by a factor $4$, since this phase undergoes the fading of both the
  giant and the intermediate galaxy populations. Nevertheless it hides
  a strong SFR drop by a factor $100$ at $z<0.2$ of the intermediate
  galaxy population. This phenomena might be combined to the local
  underdensity seen in optical redshift surveys (see \citet{zucca97},
  \citet{tresse98}), and thus the local Universe might not be an
  excellent reference local point to trace back the evolution of the
  global galaxy population.  Our observed global evolution does not
  seem to be in agreement with a continous smooth decrease from
  $z\sim2$ to $z\simeq0$ as predicted by the simulations.
\item[v)]Comparing our SFR FUV-derived densities with mid-infrared or
  H$\alpha$ SFR-derived densities, we find that at $0.4 \la z \la 2$
  the FUV is obscured by a constant factor of $\sim1.8-2$ mag, and at
  $z<0.5$ it is progressively less obscured down to $\sim0.9-1$ mag. In
  parallel, we find that from $z=0.4$ to $z=0.05$ the $B$-band
  luminosity density is more and more dominated by the early-type
  (E/Sab-like) galaxy population which is known to be dust deficient.
  Futher analysis combining VVDS-GALEX-SWIRE data will refine this
  result.
\item[vi)]We conclude that the old, most luminous and massive galaxies
  have exhausted their cold gas reservoir during their early intense
  star formation which has occured in the early Universe at $z\gg4$,
  and since $z\simeq3.5$, i.e. $\sim$ 12~Gyrs, they undergo passive
  evolution as star formation cease.  This creates excellent dry
  merger candidates, that is, in which gas has been sufficiently
  depleted that it prevents new star formation.  The small level of
  contribution of dry-mergers remains to be quantified. Younger, less
  massive and less luminous $L^{*}$ ga\-la\-xies continue to efficiently
  form stars with a large reservoir of cold gas up to $z=0.2$.  And at
  $z<0.2$ these galaxies appear to have also exhausted their gas
  supply as suggested by the $FUV$-band luminosity density noticeable
  drop, and start to evolve passively, and might be combined with a
  local underdensity.  This picture is consistent with the downsizing
  scenario for the star formation \citep{cowie96}. We recall that the
  luminosity and SFR densities are mainly dominated by the
  $L^{*}\phi^{*}$ galaxies rather than the dwarf population which is
  usually undergoing density evolution \citep{ilbert05}. At $z<3.5$
  dry mergers and morphologies evolving towards early-type galaxies
  might contribute to increase the number density of the bright
  early-types in maintaining a passive luminosity evolution as
  observed by \citet{zucca06} at $0<z<1.5$, and also to increase the
  emissivity in over-dense regions as observed by \citet{ilbert06b}.
\end{itemize}
\begin{acknowledgements}
  This research was developed within the framework of the VVDS
  consortium.  We thank the ESO staff at Paranal for their help in the
  acquisition of the data.  We thank C. Moreau for her work on the
  VVDS database at LAM.  This work was partially supported in
  France by the Institut National des Sciences de l'Univers of the
  Centre National de la Recherche Scientifique (CNRS), and its
  Programme National de Cosmologie and Programme National de Galaxies,
  and in Italy by the Ministry (MIUR) grants COFIN2000 (MM02037133)
  and COFIN2003 (num.2003020150). The VLT-VIMOS observations were 
  carried out on guaranteed time (GTO) allocated by the European
  Southern Observatory (ESO) to the VIRMOS consortium under a
  contractual agreement between the CNRS, heading a consortium of
  French and Italian institutes, and ESO to design, manufacture and
  test the VIMOS instrument.
\end{acknowledgements}
\end{document}